\documentclass[preprint]{aastex631}

\newcommand\aastex{AAS\TeX}

\accepted{for publication in The Astronomical Journal, 2024 November 19}

\shorttitle{\aastex\ Comet 289P/Blanpain and the Phoenicids}
\shortauthors{Kasuga (2024)}

\usepackage{amsmath}
\usepackage{amssymb}
\usepackage{rotating, graphicx}
\usepackage{longtable}
\usepackage[stable]{footmisc}
\usepackage{mdwlist}

\begin{document}

\title{Comet 289P/Blanpain: Near-Perihelion Activity and the Phoenicids}

\correspondingauthor{Toshihiro Kasuga}
\email{toshi.kasuga@nao.ac.jp}

\author[0000-0001-5903-7391]{Toshihiro Kasuga}
\affil{National Astronomical Observatory of Japan, 2-21-1 Osawa, Mitaka, Tokyo 181-8588, Japan}




\begin{abstract}

We present NEOWISE observations of 
Jupiter family comet 289P/Blanpain, 
the parent body of the Phoenicid meteoroid stream.   
Near-infrared images 
at 3.4$\micron$~($W1$) and 4.6$\micron$~($W2$)
were obtained near perihelion on 
two occasions: UT 2019-10-30 (inbound, heliocentric distance $R_{\rm h}$ = 1.20\,au) 
and UT 2020-01-11/12 (outbound, $R_{\rm h}$ = 1.01\,au).  
To assess faint activity, we establish constraints 
on dust production driven by the limited sublimating 
area of water ice, based on studies of the 1956 Phoenicids.   
The ejected dust mass is 
$M_{\rm d}$ = 4100\,$\pm$\,200\,kg (inbound)
and 1700\,$\pm$\,200\,kg (outbound), respectively.
The dust production rates are 
$Q_{\rm dust}$\,=\,0.01\,$-$\,0.02\,kg\,s$^{-1}$, 
corresponding to dust-to-gas production ratio 
2\,$\leqslant\,f_{\rm dg}\,\leqslant$\,6.  
The resulting fractional active area, 
$f_{\rm A}$ = 3.8$\pm$1.9$\times10^{-5}$, 
is the smallest yet reported.  
The absence of 4.6$\micron$ ($W2$) excess suggests 
that 289P contains negligible amounts of CO$_2$ and CO. 
Time-resolved analysis of weighted mean 
of $W1$ and $W2$ magnitudes 
finds a distinctive peak amplitude 
in the light curve 
having a rotational period 
$P_{\rm rot}$\,=\,8.8536\,$\pm$\,0.3860\,hr, 
however, further verification is needed.   
The perihelion-normalized nongravitational 
acceleration, $\alpha_{\rm NG}^\prime$ = 3.1$\times$10$^{-6}$, 
is approximately an order of magnitude smaller than 
the trend observed for well-studied comets, 
consistent with weak outgassing.
Current dust production from 289P, regardless 
of plausible assumptions for particle size and distribution, 
is an order of magnitude too small to produce 
the Phoenicid stream within its $\sim$300\,year 
dynamical lifetime.  
This suggests another mass supply, probably in 1743$-$1819, 
rapid rotational destruction of a sub-km precursor body,  
resulting in fragments equaling the mass 
of an object with radius $\sim$\,100\,m.  
\end{abstract}





\section{Introduction} 
\label{intro}

Comet 289P/Blanpain (hereafter 289P), 
formerly known as D/1819~W1,  
was discovered in 1819 November 
by J.-J. Blanpain \cite[][]{Kronk2003comebook}\footnote{\url{https://cometography.com/pcomets/1819w1.html}}.  
The comet was subsequently linked to 
the Phoenicid meteor shower\footnote{PHO/254 from IAU Meteor Data Center, Nomenclature \citep{JopekJenniskens11}} 
as a potential parent body due to their orbital similarities (\citeauthor{Ridley1957}\,\citeyear{Ridley1957};  
reviewed in \citeauthor{Ridley1963}\,\citeyear{Ridley1963}). 
Independent discovery and following observations 
were made for about two months until 1820 January.  
Throughout this period,  
the comet appeared small, faint and without a tail or gas.  
The orbit was determined to have a period of about 5\,years (Jupiter-family type), 
placing 289P shortly after its perihelion at the time of observations.       
Nevertheless, it remained lost for nearly two centuries 
after these initial observations.    

In 2003 November, the near-Earth asteroid (NEA) 2003~WY$_{25}$ was 
discovered by the Catalina Sky Survey \citep{Ticha2003MPEC}.    
The orbit has a semimajor axis $a$ = 3.045\,au, 
eccentricity $e$ = 0.685, 
and inclination $i$ =  $5^{\circ}.9$ 
(Epoch 2458746.5 (2019-Sep-20.0) 
from NASA JPL Small-Body Database Lookup, 
solved on 2024 July 26).  
The Tisserand parameter with respect to Jupiter, $T_J$ = 2.817, 
is consistent with dynamical classification as 
a member of the Jupiter family comets (JFCs).    
Its orbital similarity to 289P, 
coupled with backward integrations of the 2003~WY$_{25}$ orbit, 
suggested that the two bodies might be same \citep{Foglia2005IAUC8485,Micheli2005AsUAI}.

Observations and recent dynamical studies 
of the Phoenicid meteor shower revealed 
the identification of 289P with 2003~WY$_{25}$.   
On 1956 December 5, the sudden outburst 
of the shower was observed both by visual and radio-echo
in the Southern Hemisphere \cite[][]{Kronk2014meshbook}.   
Notably,  J. Nakamura witnessed 
the most significant activity of the 1956 Phoenicid outburst (300\,meteors per hour)  
on Soya, a Japanese expedition ship to the Antarctic, 
in the Indian Ocean \citep{Huruhata_Nakamura1957}.   
\cite{Watanabe2005PASJ} applied dynamical study 
to 2003~WY$_{25}$ and its orbital elements 
in the context of dust trail theory \citep{Asher2000pimo,Sato2003JIMO} 
to estimate the conditions leading to the 1956 Phoenicids outburst.  
The result found that dust trails were formed and bundled between 1743 and 1808, 
with the 1760$-$1808 trails being particularly responsible for the event, 
leading to the conclusion that 2003~WY$_{25}$ is identical to 289P 
\citep{Watanabe2005PASJ} \citep[see also,][]{JenniskensLyytinen2005,Jenniskens2006}.  

Shortly thereafter, cometary activity in 289P was confirmed to be ongoing.    
In 2004 March, \cite{Jewitt2006} optically observed 289P when 
at the heliocentric distance $R_{\rm h}$=1.64\,au, 
finding a weak coma indicative of mass-loss rate of 0.01\,kg\,s$^{-1}$ 
and a small nucleus radius ($r_{\rm n}$) of only 160\,m
\footnote{
The uncertainty is estimated several tens of percent 
based on the probable linear phase function range \citep{Jewitt2006}.  
In this study we assume the radius uncertainty of 25\%, 
corresponding to $r_{\rm n}$\,=\,160\,$\pm$\,40\,m.   
}
(an order of 
magnitude smaller than typical cometary nuclei).   
In 2013 July, an asymmetric coma and a tail were 
observed in 289P at the large $R_{\rm h}$ = 3.88\,au, 
with a maximum $V$-band apparent magnitude of 17.5 \citep{Williams2013CBET}.    
The limited mass-loss within 
the stream age ($\tau_{\rm s} \sim$ 300\,yrs) 
is insufficient to explain the estimated mass of 
the Phoenicid stream,  
suggesting that the stream might be 
produced impulsively, perhaps by breakup of the precursor body \citep[reviewed in][]{KasugaJewitt2019}. 

The short dynamical age of the Phoenicid stream 
requires a substantial dust supply from 289P.   
Since JFCs and other type of comets 
typically exhibit their strongest mass-loss activity 
around their perihelia, 
we investigate the dust production rate of 289P 
near its perihelion ($q$\,=\,0.96\,au) to assess 
its potential contribution to the Phoenicid stream.
The latest opportunity occurred from the end of 2019 
to the beginning of 2020.   
We used data acquired by 
the {\it Near-Earth Object Wide-field Infrared Survey Explorer} \citep[NEOWISE,][]{Mainzer2011ApJ, Mainzer2014ApJ} 
during both the inbound and outbound 
perihelion passages of comet 289P/Blanpain.

\section{NEOWISE}
\label{wise}

This study is based on near-infrared data obtained 
by NEOWISE \citep[][]{Mainzer2011ApJ, Mainzer2014ApJ}, 
the extended mission of WISE \citep[][]{Wright2010AJ}. 
Launched on 2009 December 14 
into a Sun-synchronous polar orbit, 
WISE was equipped with a 40\,cm telescope 
that employed beam splitters to simultaneously 
capture images at four infrared wavelength bands centered at 
3.4$\micron$ ($W1$), 4.6$\micron$ ($W2$), 
12$\micron$ ($W3$), and 22\,$\micron$ ($W4$), 
enabling a comprehensive all-sky survey. 
It started on 2010 January 7 and continued until 2010 September 29.  
As the cryogen reserve was exhausted, data acquisition terminated for $W3$ and $W4$.  
WISE further continued surveying in a post-cryogenic two-band survey mode 
until 2011 February 1, when it was placed in hibernation.
The two-band survey ($W1$ and $W2$), subsequently named NEOWISE, 
resumed 2013 December 13 and continues as the NEOWISE-R reactivation project.  

NEOWISE employs a 1024~$\times$~1024 pixel 
HgCdTe focal-plane array detector (Teledyne Imaging Sensors) 
for both the $W1$ and $W2$ bands \citep{Wright2010AJ}. 
The detector exhibits a pixel scale of $2\arcsec.75$ pix$^{-1}$ 
and a field of view (FOV) of 47$\arcmin$$\times$47$\arcmin$.
Consecutive pairs of frames, one for each band, are 
captured within the same FOV every 11\,seconds, 
with each frame having an exposure time of 7.7\,seconds. 
The NEOWISE mission data ($W1$ and $W2$ bands) are available through 
the NASA/IPAC Infrared Science Archive (IRSA\footnote{\url{https://irsa.ipac.caltech.edu/frontpage/}}).

\subsection{Data}
\label{data_screen}

We retrieved all detections of comet 289P/Blanpain using 
the IRSA search tools (GATOR\footnote{\url{https://irsa.ipac.caltech.edu/cgi-bin/Gator/nph-scan?submit=Select&projshort=WISE}} 
and WISE Image Service\footnote{\url{https://irsa.ipac.caltech.edu/applications/wise/}}).   
The queried sources in the GATOR are 
NEOWISE-R Single Exposure (L1b) \citep{https://doi.org/10.26131/irsa144}, 
WISE All-Sky Single Exposure (L1b) \citep{https://doi.org/10.26131/irsa139}, 
WISE 3-Band Cryo Single Exposure (L1b) \citep{https://doi.org/10.26131/irsa127}, 
and
WISE Post-Cryo Single Exposure (L1b) \citep{https://doi.org/10.26131/irsa124}.  
These detections were also verified using the WISE Image Service.   

For this study, we conducted the data screening process following 
the methods of the NEOWISE Comet Team \citep[][]{Stevenson2015ApJL,Rosser2018AJ,Bauer2021PSJ,Gicquel2023PSJ,Milewski2024AJ}.  
We applied the selection criteria for  
a moon angular separation $>$~30$^{\circ}$, 
and an angular distance from the nominal boundaries of 
the South Atlantic Anomaly (SAA) $>$~0$^{\circ}$.  
We selected observations having flags of \verb|cc_flags| = 0 
which ensures no contamination produced by known artifacts, 
e.g. latent images or diffraction spikes \citep{Mainzer2011ApJ}.
All scores in the profile-fit photometric quality flag (\verb|ph_qual|) 
and the frame quality (\verb|qual_frame|) were checked, 
given that cometary coma is extended and not always imaged as a point source.  
We visually inspected the single exposure frames collected above.  
An image of comet 289P was confirmed 
when its optical magnitude $< 22$\,mag, 
and data for this study were selected accordingly.   
Following \cite{Bauer2012ApJ}, 
for objects detected at multiple epochs, 
we define a ``Visit" as a set of observations 
separated by time and distinct sky regions. 
We found two observation epochs, 
Visit~$A$ (MJD~58786: UT 2019-10-30, inbound) 
and 
Visit~$B$ (MJD~58860: UT 2020-01-11/12, outbound), for 289P.

\subsection{Single Exposure Image}
\label{single}

We measured the FWHM of the 289P image in single exposure frames, 
finding values of 5$\arcsec$.8--6$\arcsec$.6 for $W1$ and 
6$\arcsec$.3--8$\arcsec$.8 for $W2$, respectively.  
These wider FWHM values, mostly beyond 
the nominal point-spread function (PSF) of 6$\arcsec$.0 
used in the GATOR photometry\footnote{\url{https://wise2.ipac.caltech.edu/docs/release/allsky/expsup/sec4_4c.html}}, 
are indicative of an extended coma surrounding 289P.  
The nominal PSF is applied for point-like objects such as asteroids and stars, 
and the aperture radius for photometry is set 7$\arcsec$.5 (=\,1.25\,$\times$FWHM).   
The short radius can not capture the entire 
extent of the emission of comets, 
leading to underestimation of the brightness of 289P. 
Therefore, to enclose the entire emission 
and improve the brightness measurements, 
we conducted photometry on the data 
selected using synthetic circular apertures projected onto the sky.  
The photometric aperture radius was set twice the FWHM in the image ($\approx12\arcsec$$\sim$$18\arcsec$) 
and the sky background was determined within a concentric 
annulus having projected inner and outer radii 
of $30\arcsec$ and $45\arcsec$, respectively.   
This aperture radius is reasonable to 
active comets in the NEOWISE/WISE data 
for enclosing most of the light source \citep[cf. $11\arcsec$-radius,][]{Stevenson2015ApJL}.  
The sky range applied here was set smaller due to the low 
Signal-to-Noise Ratio (SNR) in the data, especially for $W1$.
Images corrupted by stars, artifacts, or noise were excluded.  
Data with an extremely low SNR ($<$\,1) were rejected.  
To mitigate star contamination, particularly prevalent 
in the shorter $W1$ band, images with $W1$~SNR~$>$~$W2$~SNR 
were removed. 
The stellar thermal signatures ($\gtrsim$~2000\,K) 
have their blackbody peaks in the shorter wavelengths, 
leading to their appearances in these bands \citep{Kasuga_Masiero2022AJ}.
This issue was found in the $W1$ band from Visit~$B$ (MJD~58860: 2020-01-11/12, outbound) (Section~\ref{surfacebrightness}).   
There were no saturated images brighter than the threshold of $W1$ = 8.1\,mag, $W2$ = 6.7\,mag.  

The measured instrumental source brightness (counts in digital numbers) 
was calibrated to magnitude by referring an instrumental zero 
point magnitude\footnote{\url{https://wise2.ipac.caltech.edu/docs/release/allsky/expsup/sec4_4h.html\#CalibratedM}} 
\citep{Cutri2012wise.rept, Cutri2015nwis.rept}.  
The magnitudes were converted to the flux density (Jy) 
using the published zero points \citep{Wright2010AJ}, 
and no color corrections were applied for 289P \citep[e.g.][]{Bauer2021PSJ}.
We followed the photometry and calibration methodology 
on the IRSA website\footnote{\url{https://wise2.ipac.caltech.edu/docs/release/allsky/expsup/sec4_4h.html}} \citep{Cutri2012wise.rept, Cutri2015nwis.rept} as appropriate. 
The observation log and photometric results are shown in Table~\ref{obslog}.
The orbital information is summarized in Table~\ref{289P_orbit}.

\subsection{Composite Image}
\label{composite}

We created coadded images from both Visit~$A$ and $B$ 
using the Image Co-addition with Optional Resolution Enhancement (ICORE)
\citep[][]{Masci2013ascl}\footnote{\url{https://irsa.ipac.caltech.edu/applications/ICORE/}}.  
ICORE combines the data frames into a single averaged image,  
resamples the original NEOWISE (WISE) single-exposure frames,  
and convolves the mosaic pixel with the original PSF.  
The system extracts target object positions (RA and DEC) from individual images and 
repositions the corresponding image cutouts, assigning the recentered positions to the 
provided RA and Dec.  
The orientation of the coadded images is then adjusted to align North up and East to the left.
In this study, the image size is set at $277\arcsec$~$\times$~$277\arcsec$ 
(0.077\,deg~$\times$~0.077\,deg).
The resulting coadded images have a resampled pixel scale of $1\arcsec.0$\,pix$^{-1}$, 
down from $2\arcsec.75$\,pix$^{-1}$.

The ICORE processed both data in the Visits~$A$ and $B$ 
to create composite images.  
The $W2$ image from Visit~$B$ had a small, 
cavity-like artifact on the 289P image.  
We hesitated to conduct a cleaning process on the comet image.  
Then we alternatively applied a median-combination process using $IRAF$ 
for the Visit~$B$ dataset.    
This is because the ICORE package applies area-averaged image combine process 
(Equation~(10) in Section~7)\footnote{\url{https://irsa.ipac.caltech.edu/applications/ICORE/docs/icore.pdf}}.  
In contrast, a clipped median combination (which rejects the extreme 
pixel data values before computing the median of the remainder) can 
suppress artifacts, noise and background structures \citep{Jewitt2020ApJ}.  
This approach is effective for Visit~$B$ 
since 289P is imaged in a sky region where faint stars are densely packed.    	
Visit~$B$ obtained two single-exposure frames taken at $\Delta$ = 0.09\,au, 
both with sufficient SNR $\sim$\,25 for the combination process using $IRAF$.
The orientation of each image was rotated to bring the direction of 
the position angle to north at the top and east at the left, and 
they were shifted to align the images using fifth-order polynomial interpolation. 
The images were then median-combined into a single image, 
maintaining a pixel scale of $2\arcsec.75$ pix$^{-1}$.

\begin{figure*}[ht!]
    \begin{tabular}{cc}
       \resizebox{90mm}{!}{\epsscale{0.5} \plotone{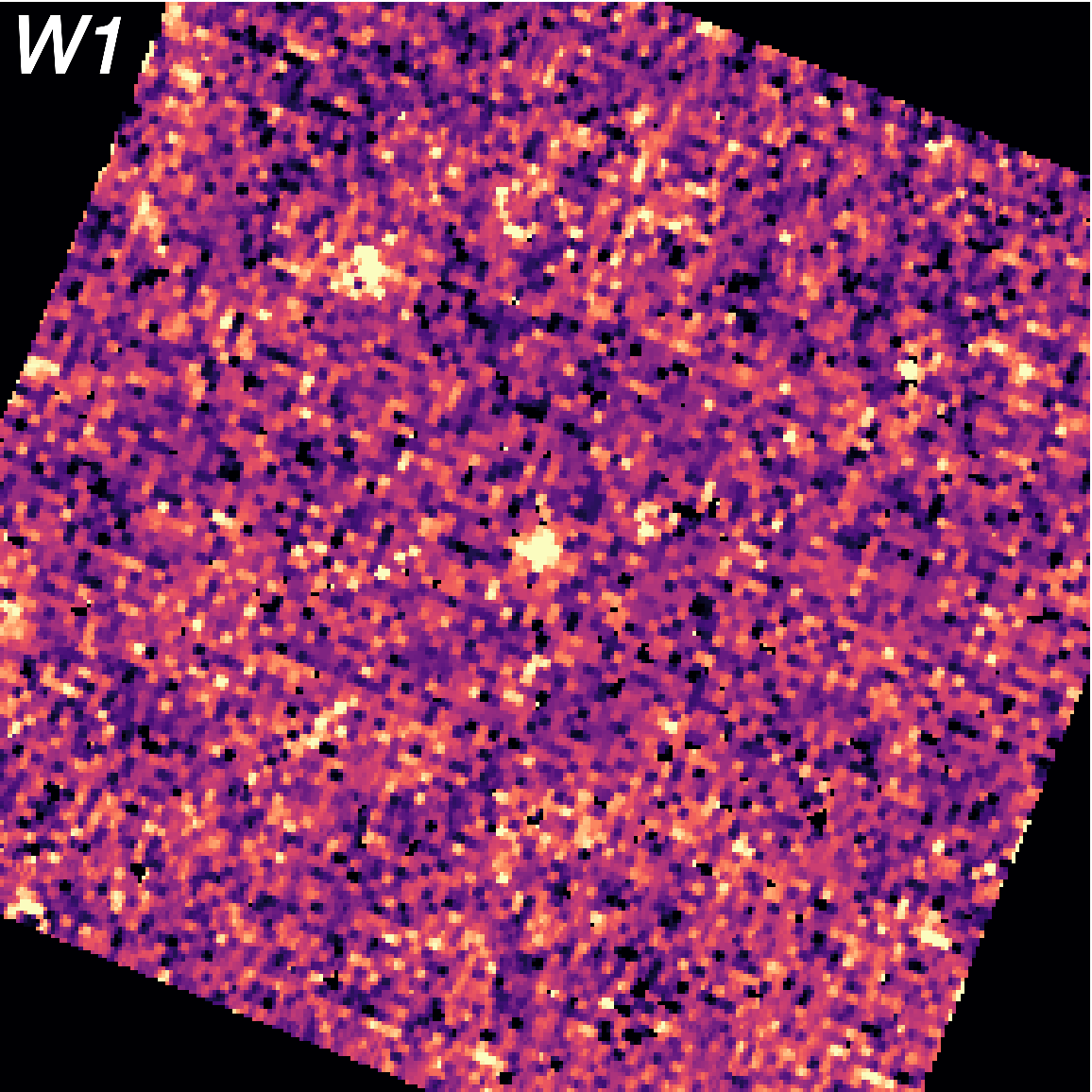}} 
       \resizebox{90mm}{!}{\epsscale{0.5} \plotone{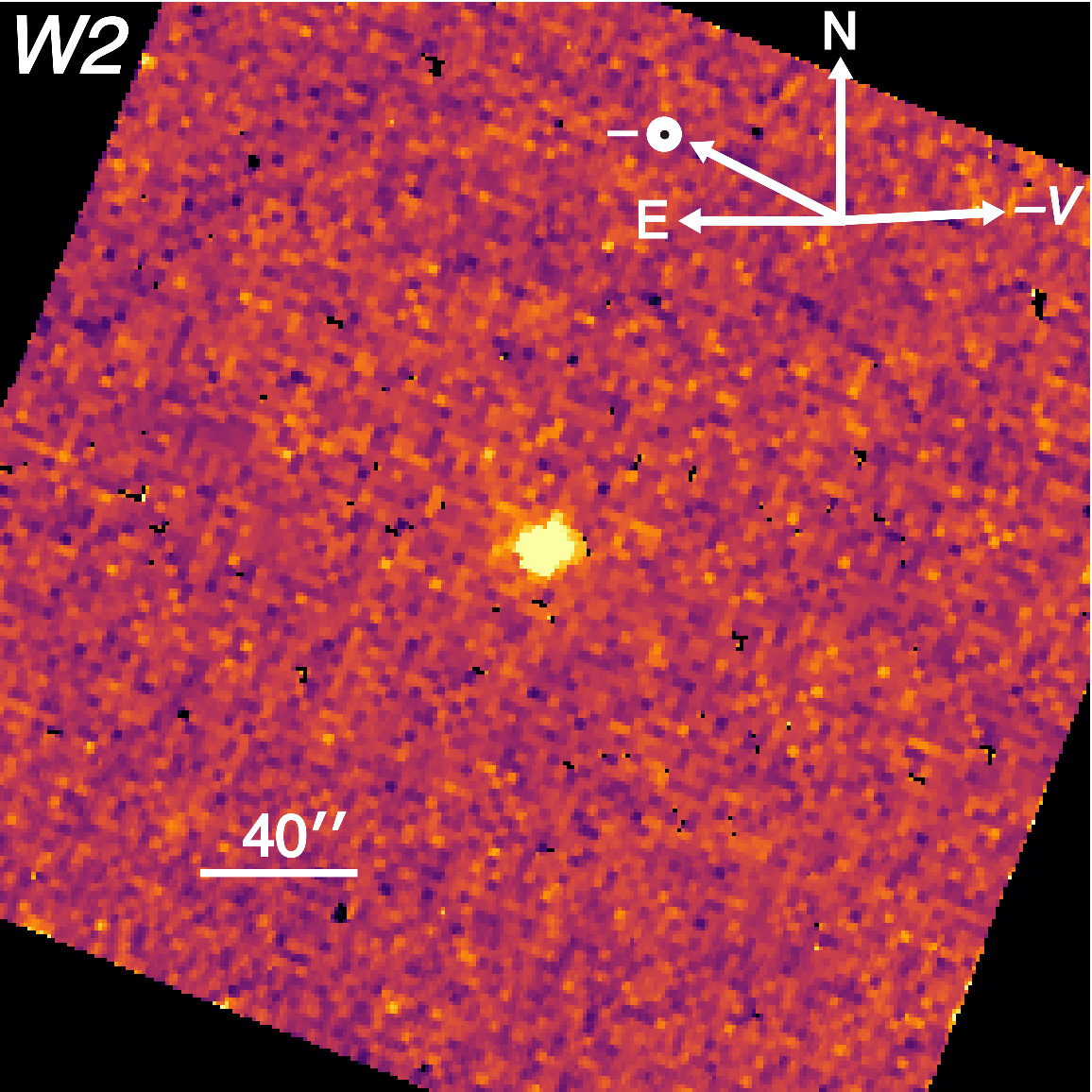}}\\
    \end{tabular}
    \caption{
    The ICORE coadded images of 289P (center) in Visit~$A$ (MJD~58786: UT 2019-10-30, inbound).  
    Both frames have a size of 277$^{\prime \prime}$~$\times$~277$^{\prime \prime}$.  
    The left panel shows the $W1$-band image (38.5\,seconds integration) with the FWHM 
    $\theta_{\rm F}$\,=\,$7\arcsec.2$, while the right panel shows
    the $W2$-band image (46.2\,seconds integration) with 
    $\theta_{\rm F}$\,=\,$7\arcsec.7$.  Heliocentric, ${\it WISE}$-centric distances and phase angle were 
    $R_{\rm h}$\,=\,1.20\,au, $\Delta$\,=\,0.40\,au 
    and $\alpha$\,=\,49.8$^{\circ}$, respectively. 
    The cardinal directions ($N$ and $E$), the direction of the 
    negative heliocentric velocity vector ($-V$), 
    and the anti-solar direction ($-\odot$) are marked.  
    A 40$\arcsec$ scale bar is also shown.  
    }
    \label{image_visitA}
\end{figure*}

\begin{figure*}[ht!]
    \begin{tabular}{ccc}
       \resizebox{75mm}{!}{\epsscale{0.5} \plotone{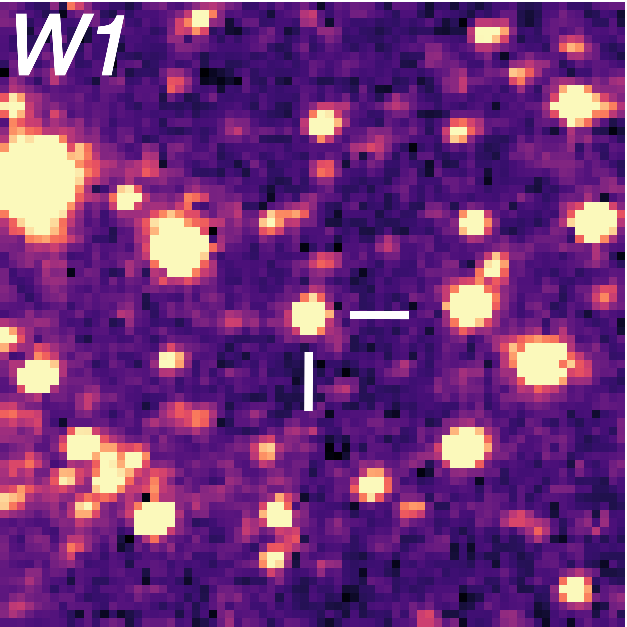}} 
       \resizebox{25mm}{!}{\epsscale{0.4} \plotone{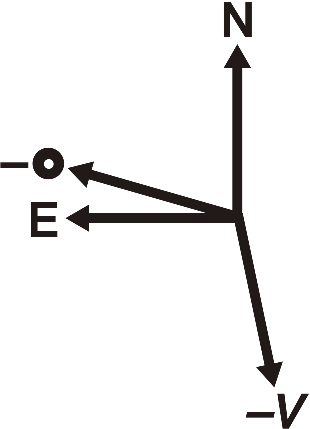}} 
       \resizebox{75mm}{!}{\epsscale{0.5} \plotone{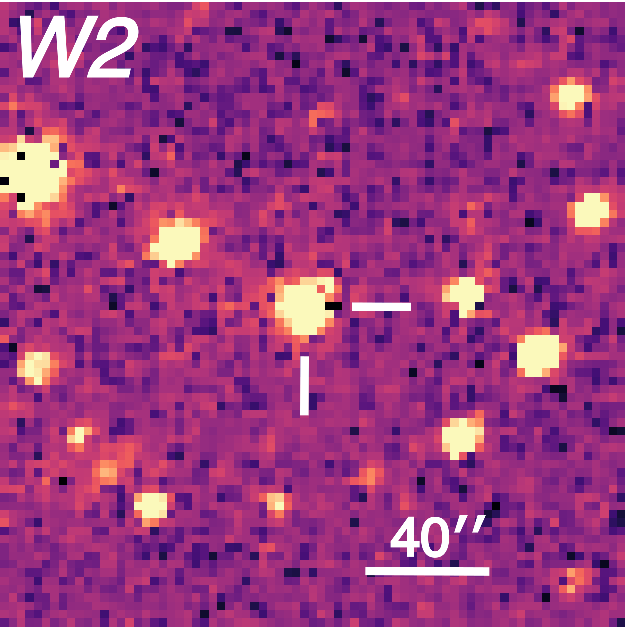}}\\
    \end{tabular}
    \caption{
    The median-combined images of 289P (center) in Visit~$B$ (MJD~58860: UT 2020-01-11/12, outbound).  
    Both frames have a size of 200$^{\prime \prime}$~$\times$~200$^{\prime \prime}$ 
    with an integration time of 15.4\,seconds.   
    The left panel shows the $W1$-band image with the FWHM $\theta_{\rm F}$\,=\,$6\arcsec.6$, 
    while the right panel shows the $W2$-band image with $\theta_{\rm F}$\,=\,$6\arcsec.9$.   
    A 40$\arcsec$ scale bar is included.  
    $R_{\rm h}$\,=\,1.01\,au, $\Delta$\,=\,0.09\,au and $\alpha$\,=\,70.2$^{\circ}$.  
    $N$ and $E$ exhibit the cardinal directions.  
    $-V$ shows the direction of the negative heliocentric velocity vector and 
    $-\odot$ shows the anti-solar direction, respectively.   
    }
    \label{image_visitB}
\end{figure*}

Figures~\ref{image_visitA} and \ref{image_visitB} 
show the ICORE coadded and median-combined images 
for Visit~$A$ and $B$, respectively.  
Photometry was conducted 
using the aperture radius twice the FWHM in the image 
($\approx12\arcsec$$\sim$$18\arcsec$) 
and the sky background was determined within 
a concentric annulus having projected inner and outer radii 
of $40\arcsec$ and $55\arcsec$, respectively.  
The results are shown in Table~\ref{mag_flux}.

\section{Analysis}
\subsection{Surface Brightness}
\label{surfacebrightness}

\begin{figure*}[htbp]
\epsscale{1} \plotone{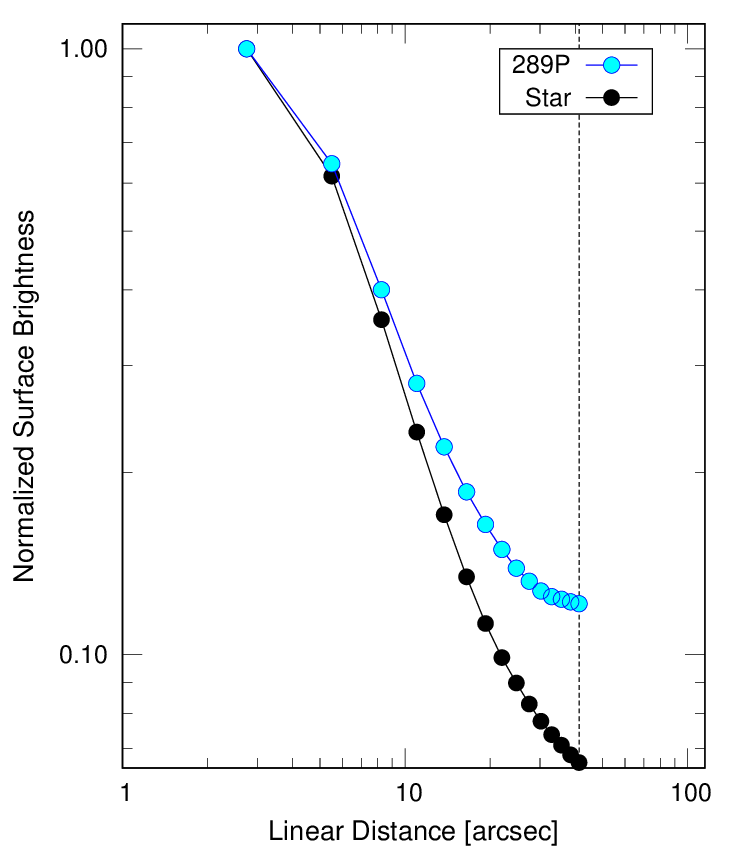} 
\caption{
Normalized surface brightness profiles of 289P (blue circles) 
and a field star (black circles), from median-combined 
$W1$ image in Visit~$B$ (MJD~58860: UT 2020-01-11/12, outbound) (Figure~\ref{image_visitB}, left panel).  
The FWHM of 289P is $\theta_{\rm F}$\,=\,$6\arcsec.6$.  
The linear distance $>$\,40$\arcsec$ is 
precluded (vertical dashed line) due to the nonuniform background, 
preventing further profile analysis.         
}
\label{289P_B_W1}
\end{figure*}

\begin{figure*}[htbp]
\epsscale{1} \plotone{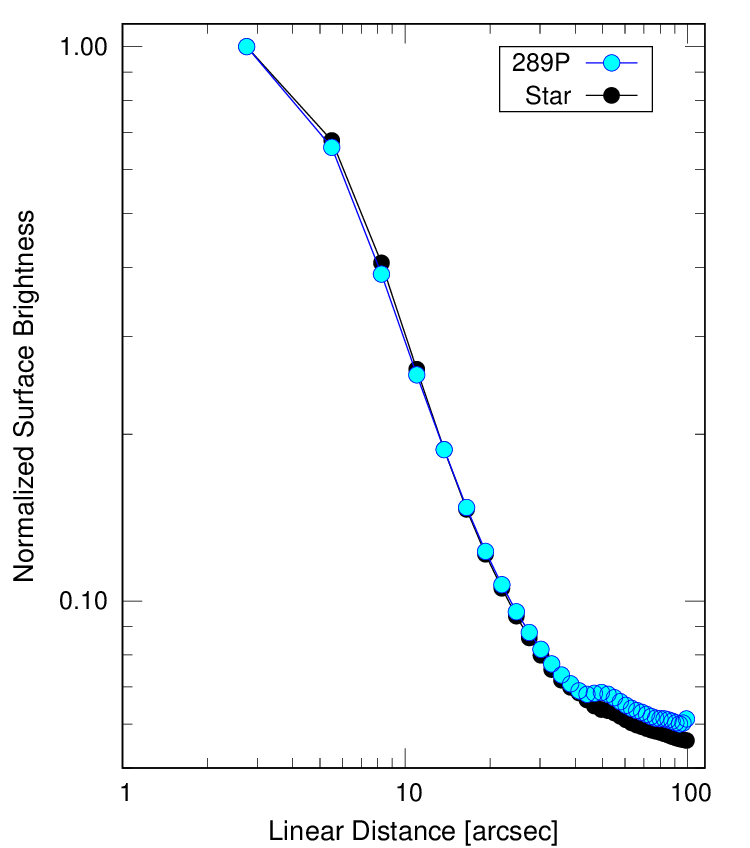} 
\caption{
Same as Figure~\ref{289P_B_W1}, but from the $W2$-band from Figure~\ref{image_visitB}, right panel.  
The FWHM of 289P is $\theta_{\rm F}$\,=\,$6\arcsec.9$.    
}
\label{289P_B_W2}
\end{figure*}

The Visit~$B$ composite images (median-combined) 
captured both 289P and nearby field stars simultaneously.  
For comparison, we measured the 
surface brightness profiles of 
both in the $W1$ and $W2$ bands.  
The surface brightness, 
as a function of the angular radius, 
was obtained using concentric annular apertures 
with a width of $2\arcsec.75$ (=\,1\,pix) 
extending to a radius of $100\arcsec$ (=\,36\,pix).  
Figures~\ref{289P_B_W1} and \ref{289P_B_W2} show 
the profiles in the $W1$ and $W2$ bands, respectively.

Figure~\ref{289P_B_W1} shows that 
the $W1$ profiles are 
broader profile compared to a field star.  
However, beyond 15\,pixels ($\sim$41$\arcsec$), 
the profile of 289P becomes severely influenced 
by the uneven structure from the background (Section~\ref{single}).  
This effect renders the data 
unsuitable for further profile analysis.  
On the other hand, 
Figure~\ref{289P_B_W2} shows that 
the $W2$ profile for 289P is 
similar to that of a field star, with a slightly broader 
extension towards the edge of the wings.  
We have a short note that these surface brightness 
profiles cannot provide constraints on the activity levels of 289P.  
As an example, the FWHM of 289P in the 
$W2$ band is $\theta_{\rm F}$=$6\arcsec.9$, 
which is $\sim$8\% larger than the NEOWISE/WISE 
angular resolution of $6\arcsec.4$ ($W2$) 
\citep[$6\arcsec.1$ for $W1$,][]{Wright2010AJ}\footnote{\url{https://wise2.ipac.caltech.edu/docs/release/prelim/}}.  
Therefore we can only compare 
the profile of 289P to those of the field stars.   

Alternatively, the profiles allow to constrain the 
speed of dust ejection from 289P.   
The profiles and visual checks of the NEOWISE data 
revealed almost no discernible coma, suggesting a compact 
and faint dust distribution in 289P \citep{Jewitt1991ASSL, Jewitt2006}.  
The weak or absence of a coma is a clear indicator of the extremely 
low velocities at which dust particles are ejected from 289P, 
as in the case of main-belt comet 133P/Elst-Pizarro \citep{Jewitt2014AJ}. 
We estimated the velocities of ejected dust, $v_{\rm d}$, 
using turnaround distance of particles toward the Sun, 
with the equation $X_{\rm R} = v_{\rm d}^2/(2\,\beta\,g_{\odot})$ \citep{Jewitt_Meech1987ApJ}.  
The dimensionless parameter $\beta$, relates to 
particle radius as $\beta$\,$\sim$\,1/$a_{\rm \mu m}$ \citep{BohrenHuffman1983}, 
where $a_{\rm \mu m}$ is the particle radius expressed in microns.   
Then we obtain a constant for dust particles ejected from the 289P nucleus \citep{Jewitt2014AJ},   
\begin{equation}
v_{\rm d}^2\,a_{\rm \mu m}  = 2\,g_{\odot}\,X_{\rm R},       
\label{turnaround}
\end{equation}
in which $g_{\odot}$ is the solar gravitational acceleration expressed 
as $g_{\odot}$ = $GM_\odot$/$R_{\rm h}^2$, 
where $G$ = 6.67 $\times$ 10$^{-11}$\,m$^3$\,kg$^{-1}$\,s$^{-2}$ is the gravitational constant, 
$M_\odot$ = 2$\times$10$^{30}$\,kg is the mass of the Sun, and 
$R_{\rm h}$ is expressed in meters.  
We apply the $W2$ composite data from Visit~$B$, 
and substitute 
$X_{\rm R} <$ 450\,km ($\theta_{\rm F}$ = $6\arcsec.9$ at $\Delta$ = 0.09\,au) 
and 
$g_{\odot}$ = 5.8$\times$10$^{-3}$\,m\,s$^{-2}$ at $R_{\rm h}$ = 1.01\,au 
to find $v_{\rm d}^2\,a_{\rm \mu m} < $~5200\,m$^{2}$\,s$^{-2}$.  
For example, with $a_{\rm \mu m}$ = 2.3, half the $W2$ wavelength \citep[][]{Bauer2011ApJ}, 
we find $v_{\rm d} <$~48\,m\,s$^{-1}$. 
For a millimeter-radius particle ($a_{\rm \mu m}$ = 1000), we find 2.3\,m\,s$^{-1}$.  
These derived dust velocities are significantly lower than those 
calculated for thermal gas outflows in active comets with the similar $R_{\rm h}$. 
As a preliminary estimate, 
the Whipple model \citep{Whipple1951} 
predicts that the $\micron$-sized particles, 
likely coupled to the outflowing gas \citep{Harmon2004comebook}, 
have speeds $v_{\rm d}$\,=\,260\,m\,s$^{-1}$ from 289P.  
Thermal gas, on the other hand, is expected to 
reach velocities $v_{\rm g}$\,=\,570\,m\,s$^{-1}$ at $R_{\rm h}$\,$\sim$\,1\,au 
with temperature $T$\,$\sim$\,278\,K \citep[Equation (10) of][]{GraykowskiJewitt19}.  
However, these calculated values are more than five times larger  
than the inferred values from the 289P profile.  
The absence of distinct coma further suggests 
low dust ejection velocities due to the weak activity of 289P.

\subsection{Sublimation from Ice Patch}
\label{patch}

The small source approximation (SSA) model \citep[][]{Jewitt2014AJ} 
provides a framework for investigating 
the mechanism responsible for 
the slow ejection of dust particles.  
We apply this model to estimate water ice sublimation 
from a small surface patch on comet 289P. 
The radius of ice sublimating patch, $r_{\rm s}$, is given by \citep[Equation (A5) of][]{Jewitt2014AJ},  
\begin{equation}
r_{\rm s} = \frac{4 \rho_{\rm d} a_{\rm d}}{3C_{\rm D}v_{\rm g}F_{\rm s}} \left( v_{\rm d}^2 + \frac{8 \pi G \rho_{\rm n} r_{\rm n}^2}{3} \right),  
\label{r_ice}
\end{equation}
where 
$\rho_{\rm d}$ is the bulk density of dust,  
$a_{\rm d}$ is radius of dust particle, 
$C_{\rm D} \sim$1 is a dimensionless drag coefficient 
which depends on the shape and porosity of the grain, 
$v_{\rm g}$ is the thermal velocity of gas molecules, 
$F_{\rm s}$ is the specific mass-loss rate of ice sublimation, 
$v_{\rm d}$ is the terminal ejection velocity of dust, 
$G$ is the gravitational constant, 
$\rho_{\rm n}$ is the bulk density of comet nucleus, 
$r_{\rm n}$ is the radius of comet 289P.  
We assume the same bulk density for dust and nucleus, 
$\rho_{\rm d}$\,=\,$\rho_{\rm n}$\,=\,1000\,kg\,m$^{-3}$, 
as used in the prior study of 289P \citep[][]{Jewitt2006}. 
The energy balance equation for an ice patch exposed at the subsolar point 
calculates $F_{\rm s}$ = 3.2$\times$10$^{-4}$\,kg\,m$^{-2}$\,s$^{-1}$ 
with $T$ = 200\,K in Visit~$A$ and 
4.7$\times$10$^{-4}$\,kg\,m$^{-2}$\,s$^{-1}$ with $T$ = 202\,K in Visit~$B$, 
respectively \citep[Equation~(6) of][]{KasugaJewitt2015}.  
Furthermore, we employ the following values,  
$v_{\rm g}\,\sim$\,480\,m\,s$^{-1}$ ($T$\,$\sim$\,200\,K at $R_{\rm h}$ = 1.01$-$1.20\,au) 
and 
$r_{\rm n}$ = 160\,$\pm$\,40\,m \citep[][]{Jewitt2006}. 
The values of $r_{\rm s}$, $a_{\rm d}$ and $v_{\rm d}$ remain unknown.  

\begin{figure*}[htbp]
\epsscale{1} \plotone{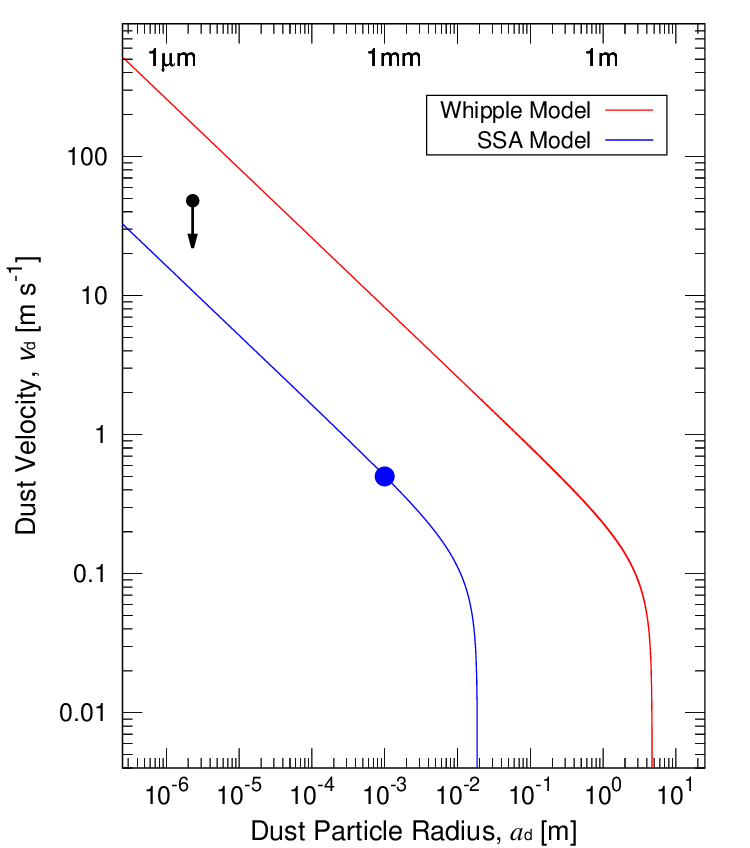} 
\caption{
Models of the ejection velocity of dust 
from 289P as a function of particle radius.  
The SSA model (blue curve) and 
the Whipple model (red curve) are 
compared.  
The blue circle marks the particle 
radius $a_{\rm d}$\,=\,1\,mm 
and the velocity $v_{\rm d}$\,=\,0.5\,m\,s$^{-1}$ 
determined by the 1956 Phoenicids 
researches (Section~\ref{patch}). 
The black symbol indicates 
$a_{\rm d}$\,=\,2.3\,$\micron$ 
and $v_{\rm d}$\,$<$\,48\,m\,s$^{-1}$ 
limited by the FWHM of 289P in the $W2$ band (Section~\ref{surfacebrightness}).  
}
\label{SSA_fig}
\end{figure*}

The 1956 Phoenicids can constrain the parameters 
of dust particle size and ejection velocity.   
The eyewitness accounts \citep[][and private communication with J. Nakamura]{Huruhata_Nakamura1957}   
indicate that the Phoenicid meteoroids had apparent 
magnitudes brighter than approximately $+$6.0\,$\sim$\,$+$6.5\,mag and 
the corresponding mass is $>$\,10$^{-6}$\,kg \citep[Table~I in][]{Lindblad1987,Kronk2014meshbook}.  
With the assumed $\rho_{\rm d}\,=\,1000$\,kg\,m$^{-3}$, 
the corresponding radius of the dust particle 
(meteoroid) is $\gtrsim$\,1\,mm.   
The dynamical study found that the 1956 Phoenicids 
primarily 
originated from the dense trails produced between 1760 and 1808, 
with slow dust ejection velocities, 0.49\,$\leqslant v_{\rm d} \leqslant$\,0.73\,m\,s$^{-1}$ 
\citep[see Table~1 in][]{Watanabe2005PASJ}.
\footnote{
The 1760$–$1808 period encompasses dust trails ejected from the parent body 
in 1760, 1766, 1771, 1776, 1782, 1787, 1792, 1797, 1803, and 1808.  
Dust particle sizes used in the simulation were insensitive to radiation pressure, 
corresponding to millimeter-scale \citep[][]{Watanabe2005PASJ}. 
}
Substituting $a_{\rm d}$\,$\sim$\,1\,mm (as the lower limit) 
and 
$v_{\rm d}$\,$\sim$\,0.5\,m\,s$^{-1}$ (median value of those computed)
into Equation~(\ref{r_ice}) 
yields 1.6\,$\pm$\,0.1m $\leqslant r_{\rm s} \leqslant$ 2.3\,$\pm$\,0.1m.   
With the determined $r_{\rm s}$ ($\approx$\,2\,m), 
the ejection velocity of dust ($v_{\rm d}$) is calculated as a function of particle radius ($a_{\rm d}$) using Equation~(\ref{r_ice}).  
The comparison of the SSA and Whipple models, both applied to 289P, 
is shown in Figure~\ref{SSA_fig}.    
The limit for dust particle of $a_{\rm d}$\,=\,2.3\,$\micron$ 
with $v_{\rm d}$\,$<$\,48 m\,s$^{-1}$, which is  
measured from the FWHM of 289P ($W2$) in 2020 (Visit~B) (Section~\ref{surfacebrightness}),  
is shown in the same Figure.  
The dust velocity is about five times 
larger than the calculated velocity by the SSA model.  
Substituting $a_{\rm d}$\,=\,2.3\,$\micron$, 
$v_{\rm d}$\,$<$\,48 m\,s$^{-1}$ and 
the same parameters above (Visit~$B$) into Equation~(\ref{r_ice}), 
we find $r_{\rm s} \lesssim$ 31\,m.  
The straightforward interpretation is that 
the ice patch area on the comet surface 
might have enlarged since 1760$-$1808.     
However, we have a note on the spatial resolution of the data.  
NEOWISE's limited resolution, 
with a pixel scale of $2\arcsec.75$ pix$^{-1}$, 
results in low angular resolution (Section~\ref{wise}).  
By comparison, for example, 
the Very Large Telescope (VLT) VISIR 
has a high resolution of $\sim$\,$0\arcsec.05$ pix$^{-1}$ 
in the mid-infrared wavelengths \citep[][]{Jewitt19Thermal}.   
The resolution gap is 
180\,km\,pix$^{-1}$ (NEOWISE) versus 3\,km\,pix$^{-1}$ (VLT) 
at the distance of 289P in Visit~$B$ ($\Delta$=0.09\,au).  
Thus, the measured dust velocity by NEOWISE ($v_{\rm d}$\,$<$\,48 m\,s$^{-1}$)  
would be an overestimate and should be considered an upper limit.  
The corresponding radius of ice patch ($r_{\rm s} \lesssim$ 31\,m) is considered likewise.  
Higher spatial resolution observations 
would improve estimates of dust velocity and ice patch area.  
Such data would allow us to constrain 
the long-term stability of 289P's surface ice source regions 
and its overall activity.

The radius of ice sublimating patch, $r_{\rm s}$, 
is related to the critical radius of dust particles 
to be launched, $a_{\rm c}$, using 
Equation~(A6) in \cite{Jewitt2014AJ},  
\begin{equation}
a_{\rm c}~ \lesssim~\frac{9 C_{\rm D} v_{\rm g} F_{\rm s} r_{\rm s}}{32 \pi G \rho_{\rm n} \rho_{\rm d} r_{\rm n}^2}.
\label{ac}
\end{equation}
Substituting the above parameters, 
we find that the critical radius of dust particle to be ejected is  
$a_{\rm c} \lesssim$\,2\,cm.  
This size is consistent with relatively large meteoroids 
which are typically seen as fireballs \citep[][]{Lindblad1987}.  
Actually, J. Nakamura observed bright, fireball-class 
Phoenicids too during its peak activity period \citep{Huruhata_Nakamura1957}, 
as noted by \cite{Watanabe2005PASJ}.
The estimated critical particle size is consistent 
with the observational result of the Phoenicids.

\subsection{Rotation Period}
\label{P_rotation}

\begin{figure*}[ht!]
    \begin{tabular}{cc}
       \resizebox{95mm}{!}{\epsscale{0.5} \plotone{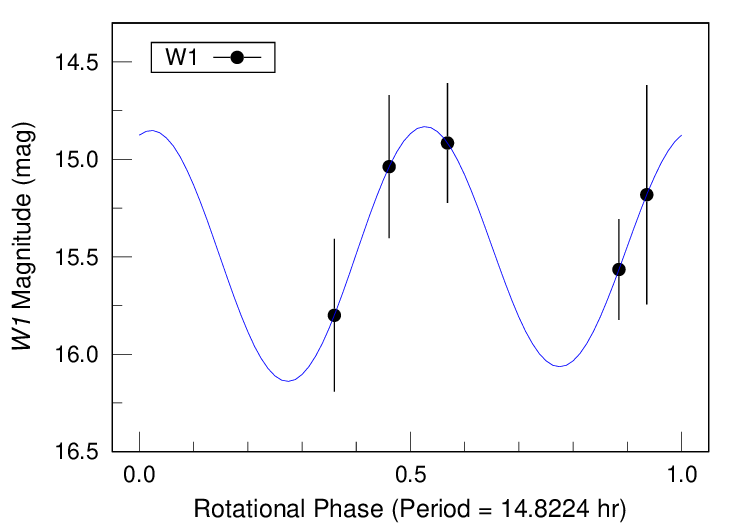}} 
       \resizebox{95mm}{!}{\epsscale{0.5} \plotone{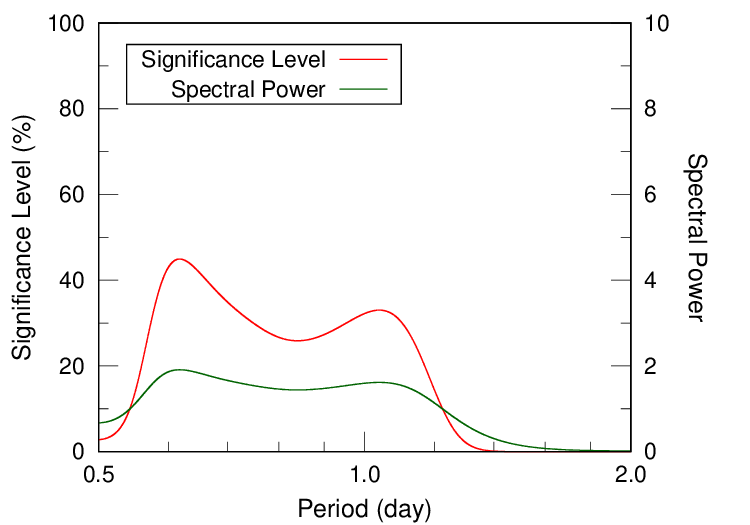}}\\
    \end{tabular}
    \caption{
    The left panel shows $W1$-band photometry of 289P 
    in Visit~$A$ (MJD~58786: UT 2019-10-30, inbound), 
    phased to the two-peaked period 
    $P_{\rm rot}$\,=\,14.8224\,$\pm$\,2.9899\,hr~$\approx$~15\,$\pm$\,3\,hr.
    The blue curve displays fitting result having amplitude $\sim$\,1.3\,mag.  
    The right panel shows spectral analysis curves for the data.  
    The maxima at $P_{\rm rot}$\,=\,0.6176\,day (=\,14.8224\,hr)
    is taken as the best solution.    
    The significance level is $\sim$45\%.  
     }
    \label{Prot_W1}
\end{figure*}

\begin{figure*}[ht!]
    \begin{tabular}{cc}
       \resizebox{95mm}{!}{\epsscale{0.5} \plotone{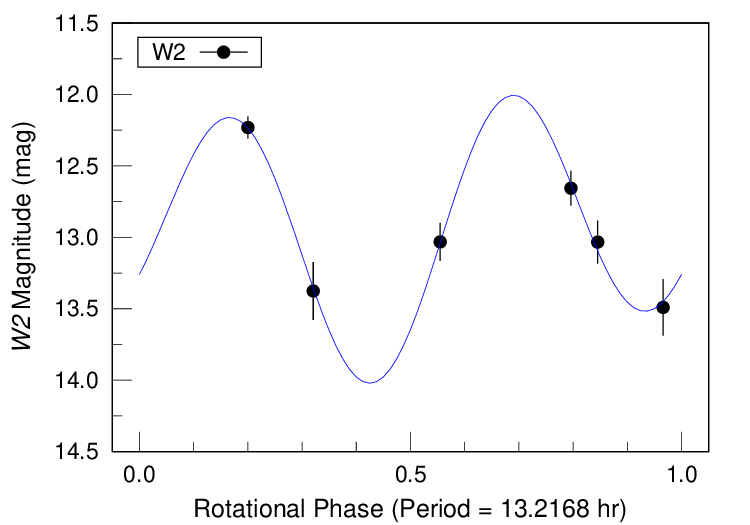}} 
       \resizebox{95mm}{!}{\epsscale{0.5} \plotone{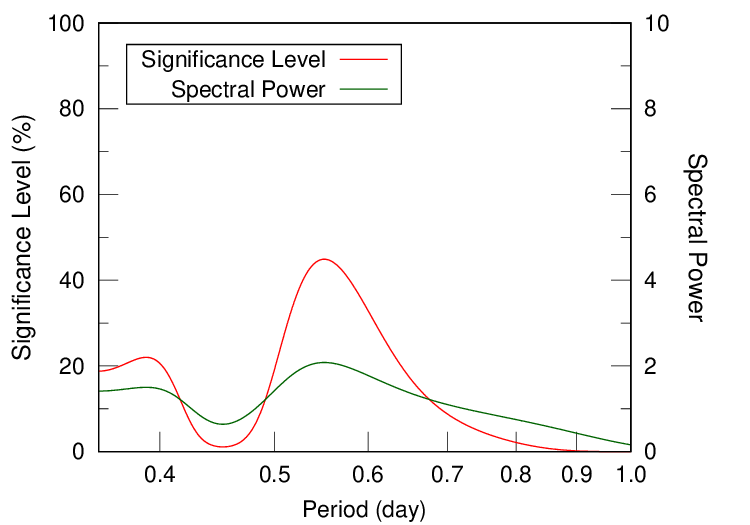}}\\
    \end{tabular}
    \caption{Same as Figure~\ref{Prot_W1} but from the $W2$-band, presenting 
    the $P_{\rm rot}$ =13.2168$\pm$2.3551\,hr~$\approx$~13\,$\pm$\,2\,hr 
    with an amplitude $\sim$2.0\,mag in the left panel  
    and the maxima $P_{\rm rot}$\,=\,0.5507\,day (=\,13.2168\,hr) in the right panel.  
    The significance level is $\sim$45\%.  
     }
    \label{Prot_W2}
\end{figure*}

Comet 289P/Blanpain likely exhibits weak activity driven by ice sublimation.  
While it is very subtle, the outgassing might corrupt 
the strict periodicity of the nucleus light curve \citep{Jewitt1991ASSL}.    
Nevertheless, Visit~$A$ provides consecutive datasets of brightness 
to search for the rotation period of 289P (Table~\ref{obslog}).   
We applied spectral analysis using 
the Discrete Fourier Transform (DFT) algorithm \citep[][]{Lomb1976ApSSL,Scargle1982ApJ}  
to the time-series data acquired in the $W1$ and $W2$, respectively.  
This method employs a significance level to assess 
the quality of the fit at a given period, using 
spectral power as a function of angular frequency. 
The highest significance level corresponds to 
maximum power at the frequency, 
indicating the most convincing solution for the periodicity. 
The light curve was assumed to be a double-peaked shape, 
as commonly observed in elongated small bodies 
in the solar system.  
Then we searched the rotational period, $P_{\rm rot}$, 
using the uncertainty estimation equation given by 
\cite{GillilandFisher1985PASP} \citep[Equation~(7) in][]{KasugaJewitt2015} 
and we found 
$P_{\rm rot}$\,=\,15\,$\pm$\,3\,hr in the $W1$ band 
and 
$P_{\rm rot}$\,=\,13\,$\pm$\,2\,hr in the $W2$ band, respectively.  
The phased light curve for each band is 
shown in Figures~\ref{Prot_W1} and \ref{Prot_W2}. 
The $W1$ light curve (Figure~\ref{Prot_W1}) 
shows low SNR and unclear systematic variations, 
with large photometric uncertainties relative to its amplitude range.   
In contrast, the $W2$ light curve (Figure~\ref{Prot_W2}) shows moderate SNR 
but shows an inconsistent shape compared to the $W1$ light curve. 
Both rotational periods are determined with a significance level of $\sim$45\%.

\begin{figure*}[ht!]
    \begin{tabular}{cc}
       \resizebox{95mm}{!}{\epsscale{0.5} \plotone{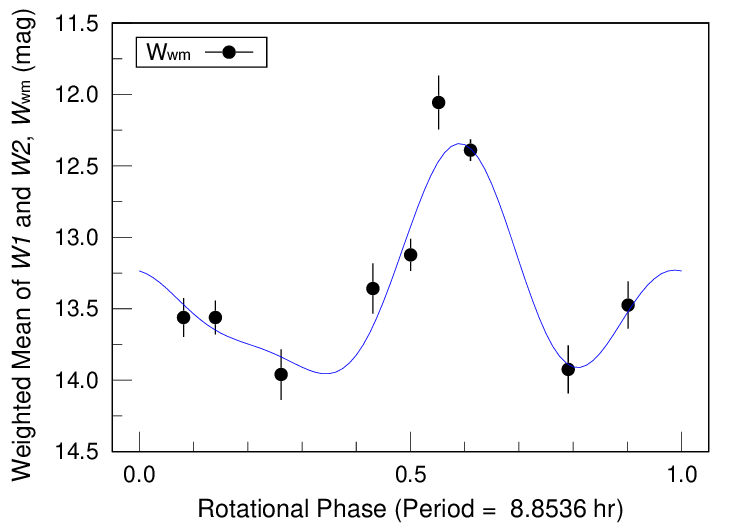}} 
       \resizebox{95mm}{!}{\epsscale{0.5} \plotone{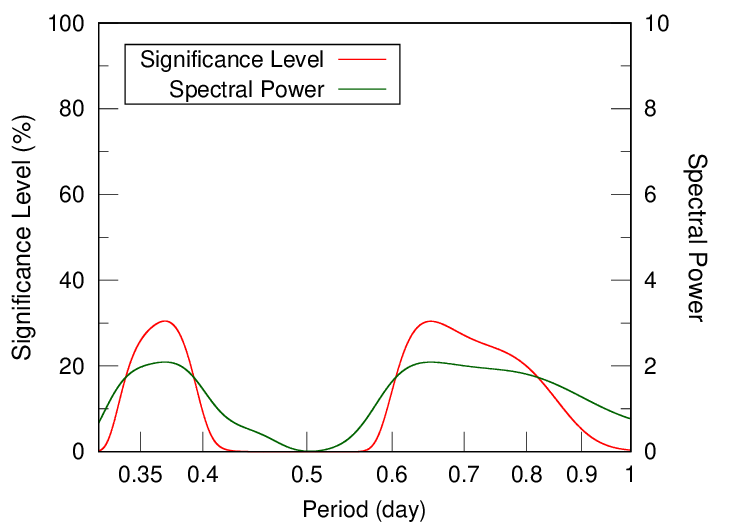}}\\
    \end{tabular}
    \caption{
    Same as Figure~\ref{Prot_W1} but from the weighted mean of $W1$ and $W2$ magnitudes ($W_{\rm wm}$), 
    presenting the $P_{\rm rot}$\,=\,8.8536 $\pm$ 0.3860\,hr with an amplitude $\sim$1.6\,mag in the left panel  
    and the maxima $P_{\rm rot}$\,=\,0.3689 \,day (=\,8.8536\,hr) with a significance level of 30.5\,\% in the right panel. 
    Another candidate $P_{\rm rot}$\,$\sim$\,15.6\,hr ($\sim$0.65\,day) has a significance level of 30.4\,\%.  
     }
    \label{Prot_W12}
\end{figure*}

To find a better constraint on the rotational period, 
we used a weighted mean of 
the $W1$ and $W2$ magnitudes. 
Missing data points in Table~\ref{obslog} 
were interpolated and extrapolated, 
assuming uncertainties of 0.3\,mag and 0.2\,mag 
for $W1$ and $W2$, respectively.  
The calculated weighted mean, $W_{\rm wm}$, 
is presented in Table~\ref{lightcurve_log}.  
By fitting $W_{\rm wm}$, 
we identify a primary rotational period,  
$P_{\rm rot}$\,=\,8.8536 $\pm$ 0.3860\,hr,  
with a significance level of 30.5\,\%.
The phased light curve and significance level 
are shown in Figure~\ref{Prot_W12}. 
A secondary candidate period,  
$P_{\rm rot}$\,$\sim$\,15.6\,hr ($\sim$0.65\,day),  
with a similar significance level of 30.4\,\%, 
was also identified, consistent with individual band analysis. 
Light curves with periods around $P_{\rm rot}$\,$\sim$\,13\,$-$\,15\,hr 
exhibit a typical double-peak pattern 
with similar amplitudes (Figures~\ref{Prot_W1} and \ref{Prot_W2}).  
However, Figure~\ref{Prot_W12} (left panel) 
reveals a distinctive peak in the amplitude 
curve at a phase $\sim$\,0.6.  
This peak might indicate an active 
spot on the 289P surface.  
In some observed active comets, 
the rotation period of the nucleus 
can be inferred from periodic structures,  
like jets and spirals (localized activities),  
in the coma, even when the nucleus 
is too faint for photometric isolation 
\citep[][]{SamarasinhaAHearn1991Icar,Samarasinha2004comebook}, as noted by \cite{Jewitt2021AJTorques}.  
From this perspective,  
we adopt $P_{\rm rot}$\,=\,8.8536 $\pm$ 0.3860\,hr 
as the principal rotational period of 289P.

We have a note that 
the derived low significance levels of $\sim$30$-$45\,\% 
indicate high uncertainties in the determined periods.  
For comparison, the rotational periods of the dormant JFC 
169P/NEAT \citep[Figure~3 in][]{Kasuga_et_al2010} 
and NEA 2003~EH$_1$ \citep{KasugaJewitt2015}
were determined with a significance level exceeding $\gtrsim$\,99\,\%, 
corresponding to the maximum power in the spectral analysis. 
The resulting uncertainties of these periods were 
$\lesssim$\,0.01\,hr, more than an
order of magnitude smaller than those of 289P.   
Thus, the obtained light curves for 289P may not be entirely reliable 
due to the low SNR and the limited number of data points.

On the contrary, the fitted light curve models may 
indicate the presence of an expanded coma.   
Assuming the variations originated solely from 
the nucleus rotation, the maximum photometric 
range would be 
$\Delta m_{\rm w}$ = 1.3\,$\sim$\,2.0 (Figures~\ref{Prot_W1}, \ref{Prot_W2}, and \ref{Prot_W12}).  
This value can be used to derive a lower limit to 
the intrinsic axis ratio, ${\it b/c}$, 
between long axis ${\it b}$ and short axis ${\it c}$.
Assuming its perpendicular rotation
(the rotational axis is perpendicular to our line of sight), 
the equation, $b/c$\,=\,$10^{0.4 \Delta m_{\rm w}}$,  
gives the ratio $b/c$ $\sim$ 
3.3\,$-$\,6.3.  
This derived intrinsic axis ratio suggests a highly elongated shape, 
far exceeding those observed in typical km-sized cometary nuclei 
\citep[${\it b/c}$\,$\ge$\,1.5;][]{Jewitt2004} \citep[see also,][]{Kokotanekova2017,Knight2023arXiv}.  
Substituting $b/c$ $\sim$ 3.3\,$-$\,6.3 and $P_{\rm rot}$\,$\sim$\,8.8536\,hr 
into Equation~(4) of \cite{JewittLi10}, 
we find the critical bulk density 1500\,$-$\,5500\,kg\,m$^{-3}$.    
This uncertainty is extremely large.   
Given the current data, further interpretation is not justified 
due to probable contamination from the coma.

\subsection{Flux Model: Reflected Sunlight and Thermal Emission }
\label{model}

Generally, active comets detected by NEOWISE are composed of  
reflected sunlight, gas (CO$_2$ and CO) and thermal emissions.  
Reflected sunlight is observed in the $W1$ band,  
attributed to light scattering by cometary dust particles.   
The gas and thermal emissions are observed in the $W2$ band 
and at longer wavelengths \citep{Bauer2015ApJ}.  
However, the presence of CO$_2$ and CO 
is not confirmed in 289P (discussed in Section~\ref{W2Q}).  
In this section, we present the models for 289P 
applied to reflected sunlight and thermal emission.  

\begin{figure*}[ht!]
\epsscale{1} \plotone{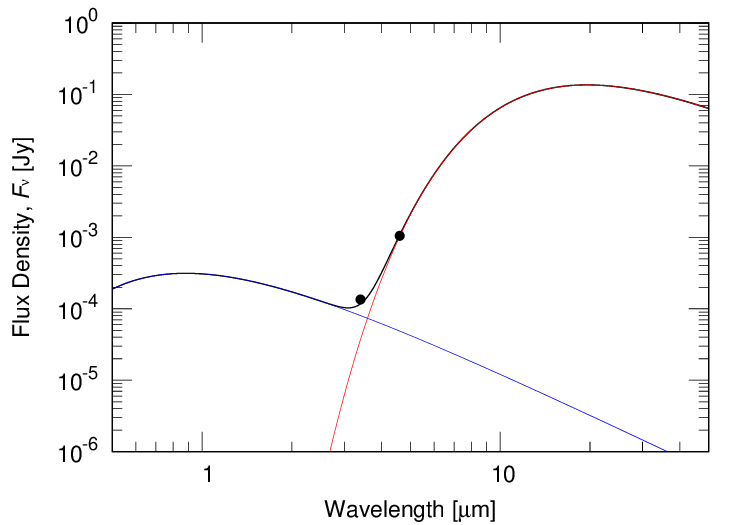} 
\caption{
Calculated spectral energy distribution and measured flux densities 
from the 289P composite image in Visit~$A$ (MJD~58786: UT 2019-10-30, inbound).   
The flux densities at $W1$~(3.4$\micron$) 
and $W2$~(4.6$\micron$) are shown as points.  
The uncertainties are within the point size. 
The reflected sunlight model (blue line), thermal model (red line) 
and combined signal (black line) are over-plotted.   
The geometry, $R_{\rm h}$=1.20\,au, $\Delta$=0.40\,au, and $\alpha$=49.8$^{\circ}$, are used for the models.  
}
\label{289P_A}
\end{figure*}

\begin{figure*}[ht!]
\epsscale{1} \plotone{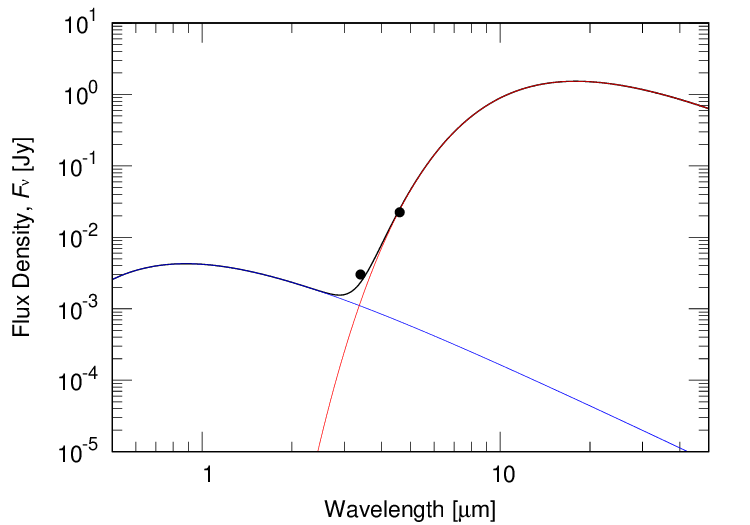} 
\caption{
Same as Figure~\ref{289P_A} but in Visit~$B$ (MJD~58860: UT 2020-01-11/12, outbound).  
$R_{\rm h}$=1.01\,au, $\Delta$=0.09\,au, and $\alpha$=70.2$^{\circ}$ are used for the models.
}
\label{289P_B}
\end{figure*}

For active comets with comae,  
the flux density, $F_\nu$, is a combination of reflected sunlight, $F^{\rm Ref}_\nu$, 
and thermal emission, $F^{\rm IR}_\nu$, primarily from dust particles.  
The equation is expressed as 
\begin{equation}
F_\nu  =  F^{\rm Ref}_\nu + F^{\rm IR}_\nu.  
\label{F_nu}
\end{equation}
In which $F^{\rm Ref}_\nu$~(Jy) is calculated by correcting 
the Sun-observer-object distance using the equation  
\begin{equation}
F^{\rm Ref}_\nu   =  \frac{\pi\,B_\nu (T_\odot)\,R_\odot^2}{R_{\rm h}^2}\,A\,\pi\,\left(\frac{D_{\rm e}}{2}\right)^2\,\frac{\Phi_{\rm HM}(\alpha)}{\Delta^2},  
\label{Fvis}
\end{equation}
where $B_\nu$ is the Planck function (Jy~sr$^{-1}$) at the Solar temperature $T_\odot$ = 5778\,K, 
$R_\odot$ = 6.957 $\times$10$^{10}$~cm is the Solar radius, 
$R_{\rm h}$ is heliocentric distance (cm).  
The $A \approx A_{\rm v}$ = $p_{\rm v}\,q_{\rm p}$ 
is the bolometric Bond albedo, where 
$p_{\rm v}$ is the visible geometric albedo 
and $q_{\rm p}$ is the phase integral. 
We assume $p_{\rm v}$=0.04 as typical of the nuclei 
of short-period comets with low-$T_{\rm J}$ \citep{Lamy2004CometII,Fernandez2013Icar} 
and $q_{\rm p}$ = 0.30, based on measurements 
for the short-period JFCs by telescopes \citep{Fernandez2003AJ,Buratti2004Icar} 
and spacecraft \citep[$q_{\rm p} \sim$ 0.20$-$0.30,][]{Knight2023arXiv}.    
$D_{\rm e}$ is the effective diameter of a circle whose area 
is equal to the combined area of all the dust particles and nucleus (cm).  
$\Phi_{\rm HM}(\alpha)$ is the Halley-Marcus composite phase 
function for scattered sunlight by dust particles in cometary comae 
\cite[][]{Schleicher1998Icar,Marcus2007ICQ_29_39M} in which  
$\alpha$ is the phase angle (in degrees),  
and 
$\Delta$ is the ${\it WISE}$-centric distance (cm).  
In Equation~(\ref{F_nu}), $F^{\rm IR}_\nu$~(Jy) is calculated by  
\begin{equation}
F^{\rm IR}_\nu =  \epsilon \left(\frac{D_{\rm e}}{2}\right)^2 \frac{B_\nu (T_{\rm d})}{\Delta^2}
\label{F_thermal}
\end{equation}
where $\epsilon$ = 0.9 is the infrared emissivity for the typical value measured from silicate powders \citep{HovisCallahan1966,Lebofsky1986Icar},
$B_\nu (T_{\rm d})$ is the Planck function (Jy~sr$^{-1}$) at the dust temperature, $T_{\rm d}$ (Kelvin).    
$D_{\rm e}$ (cm) and $\Delta$ (cm) are the same in Equation~(\ref{Fvis}).  
The cometary dust temperature, $T_{\rm d}$,  
is empirically adopted as approximately 
3\% warmer than the blackbody temperature, 
$T_{\rm bb}$ = 278$\times$$R_{\rm h}^{-1/2}$ (kelvin), 
yielding $T_{\rm d}$\,$\approx$\,1.03$\times$$T_{\rm bb}$ \citep[][]{Stevenson2015ApJL}.
These models were applied to the composite image measurements (Table~\ref{mag_flux}).  
As for the observing geometry, 
we used those of median values in Table~\ref{obslog}: 
$R_{\rm h}$\,=\,1.20\,au, $\Delta$\,=\,0.40\,au and $\alpha$\,=\,49.8$^{\circ}$ for Visit~$A$,  
and     
$R_{\rm h}$\,=\,1.01\,au, $\Delta$\,=\,0.09\,au and $\alpha$\,=\,70.2$^{\circ}$  for Visit~$B$, 
respectively.   
Using Equation~(\ref{F_nu}), we conducted least-square fits 
to the measured flux densities in Table~\ref{mag_flux}: 
$W1$ = 1.35$\pm$0.17$\times$10$^{-4}$\,Jy 
and 
$W2$ = 1.05$\pm$0.06$\times$10$^{-3}$\,Jy for Visit~$A$, 
and     
$W1$ = 3.02$\pm$0.13$\times$10$^{-3}$\,Jy 
and 
$W2$ =  2.24$\pm$0.08$\times$10$^{-2}$\,Jy for Visit~$B$, 
respectively.      
Then, we obtain $D_{\rm e}$.
The results are shown in Table~\ref{final_results}. 
The modeled spectral energy distributions and measured flux densities 
for Visits~$A$ and $B$ 
are shown in Figures~\ref{289P_A} and \ref{289P_B}, respectively. 
The obtained values,  
$D_{\rm e}$ = 1.43$\pm$0.04\,km for Visit~$A$ ($R_{\rm h}$=1.20\,au)
and 
$D_{\rm e}$ = 0.96$\pm$0.01\,km for Visit~$B$ ($R_{\rm h}$=1.01\,au),  
are consistent within a factor of two, 
indicating the stable near-perihelion activity during both Visits (confirmed in Section~\ref{W1Q}).  

We have a short technical note on model fitting.  
To ensure a meaningful least-square fit, 
we employed a single free parameter ($D_{\rm e}$) in the model.  
Including geometric albedo ($p_{\rm v}$) as an additional 
free parameter would create an overdetermined system 
with exactly two observed data points ($W1$ and $W2$) 
and 
two free parameters  ($D_{\rm e}$ and $p_{\rm v}$).  
This leads to zero degrees of freedom (\,2\,$-$\,2 = 0), 
potentially underestimating the uncertainties 
in the derived parameters. 
To estimate dust production rates in the next section, 
$D_{\rm e}$ and its uncertainty are focused.

\section{Results}

The NEOWISE images present an indistinct coma in 289P.   
Active comets can be analyzed using 
the signals in the $W1$ and $W2$ bands 
to estimate the dust production rate and 
production rates of gases like CO$_2$ and CO,  
as demonstrated in previous studies 
\citep{Bauer2008PASP,Bauer2011ApJ,Bauer2015ApJ,Bauer2021PSJ}. 
In this section, we derive the dust production rate from 289P in Visits~$A$ and $B$, respectively, 
and explore the potential reasons for the lack of detectable gas emission.  

\subsection{Dust Production Rate: $W1$}
\label{W1Q}

We estimate the dust production rate using two methods.     
One uses the scattering cross-section of 
dust particles from coma in 289P.  
Another uses the $Af\rho$-method \citep{AHearn1984}, 
similar to methods applied to comets observed by NEOWISE \citep{Bauer2008PASP,Bauer2011ApJ,Bauer2021PSJ,Milewski2024AJ}. 
These two results are compared for verification.  

The cross-section of all the dust particles, $C_{\rm d}$, is 
obtained by subtracting the nucleus cross-section of 289P, $C_{\rm n}$,   
from the total cross-section, $C_{\rm e}$ (which includes contributions 
both from dust and nucleus):  $C_{\rm d}$ = $C_{\rm e}$ $-$ $C_{\rm n}$.  
The total cross-section ($C_{\rm e}$) and nucleus cross-section ($C_{\rm n}$) 
are calculated using 
$C_{\rm e}$ = $\pi$$D_{\rm e}^2$/4 (where $D_{\rm e}$ is from Table~\ref{final_results}) 
and $C_{\rm n}$ = $\pi$$r_{\rm n}^2$ 
(where $r_{\rm n}$\,=\,160\,$\pm$\,40\,m), 
respectively.   
Then the ejected dust mass, $M_{\rm d}$, is estimated by
\begin{equation}
M_{\rm d} = \frac{4}{3}\,a_{\rm d}\,\rho_{\rm d}\,C_{\rm d}, 
\label{M_d}
\end{equation}
where $a_{\rm d}$\,=\,2.0\,$\micron$ \citep[half the median of 
the $W1$ and $W2$ wavelengths, adjusted from][]{Bauer2011ApJ} 
is the radius of the dust particle contributing to dust coma brightness 
and $\rho_{\rm d}$ \,=\,1000~kg\,m$^{-3}$ is the assumed bulk density.   
For the particle radius modification ($a_{\rm d}$\,=\,2.0\,$\micron$), 
this is because $D_{\rm e}$ was determined by 
both $W1$ and $W2$ bands (Section~\ref{model}), 
resulting in a slight adjustment to the particle radius 
from the originally applied value of 1.7\,$\micron$.  
The residence time, $\tau$, within the photometry aperture 
is estimated by $\tau$=$\rho$/$v_{\rm d}$, 
where $\rho$ is the radius of 
the aperture projected to 
the distance of the comet (=\,2\,$\times$\,FWHM at $\Delta$ in Table~\ref{obslog}) 
and $v_{\rm d}$= 11.5\,m\,s$^{-1}$ (for $a_{\rm d}$ = 2.0\,$\micron$) 
is the dust ejection velocity calculated from Equation~(\ref{r_ice}).  
The dust production rate, $Q_{\rm dust}$ (kg s$^{-1}$), is given by 
\begin{equation}
Q_{\rm dust} = \frac{M_d}{\tau} =  \frac{4\,a_{\rm d}\,\rho_{\rm d}\,C_{\rm d}\,v_{\rm d}}{3\,\rho}.  
\label{Qdust}
\end{equation}
The results ($Q_{\rm dust}$), including $C_{\rm d}$ and $M_{\rm d}$, are 
shown in Table~\ref{final_results}.  
We obtain $Q_{\rm dust}$ = 1.0$\pm$0.1$\times$10$^{-2}$\,kg\,s$^{-1}$ in Visit~$A$ ($R_{\rm h}$ = 1.20\,au) 
and $Q_{\rm dust}$ = 2.0$\pm$0.3$\times$10$^{-2}$ in Visit~$B$ ($R_{\rm h}$ = 1.01\,au), respectively, 
as it orbits around perihelion ($q$\,=\,0.96\,au).  
This low-level activity is consistent with 
the previous optical observation, $Q_{\rm dust}$$\sim$0.01\,kg\,s$^{-1}$ 
at $R_{\rm h}$\,=\,1.64\,au in 2004 \citep{Jewitt2006}.

Next, we applied the $Af\rho$-method \citep{AHearn1984}.  
The parameter, $Af\rho$ (cm), is given by  
\begin{eqnarray}
Af\rho & = & \frac{A(\alpha) f \rho}{\Phi_{\rm HM}(\alpha)} \nonumber \\
           & = &  \frac{(2 R_{\rm au} \Delta)^2}{\Phi_{\rm HM}(\alpha)\,\rho}~10^{(W1_\odot - W1)}, 
\label{Afro}
\end{eqnarray}
where $A$ is the phase angle($\alpha$)-corrected albedo of the dust 
which equals to $A(\alpha)$/$\Phi_{\rm HM}(\alpha)$ \citep{Blaauw2014MPS}, 
$f$ is the filling factor of the dust, 
$\Phi_{\rm HM}(\alpha)$ is again the Halley-Marcus composite phase function, 
$\rho$ is the radius of the aperture projected to the distance of the comet (the same above), 
$R_{\rm au}$ is the heliocentric distance expressed in au, 
$\Delta$ is the WISE-centric distance in cm, 
$W1$ is the measured magnitude at $W1$ of 289P from the composite image, 
and $W1_\odot$ is the solar magnitude in the $W1$ band.   
Substituting the $W1$ values from Table~\ref{mag_flux} 
(15.903$\pm$0.133\,mag in Visit~$A$ and 12.525$\pm$0.046\,mag in Visit~$B$, respectively) 
and 
$W1_\odot$ = $-$28.31\,mag\footnote{\url{https://mips.as.arizona.edu/~cnaw/sun.html}} 
into Equation~(\ref{Afro}),  
we find $Af\rho$ = 2.9$\pm$0.4\,cm in Visit~$A$ 
and $Af\rho$ = 10.6$\pm$0.4\,cm in Visit~$B$, respectively. 
These $Af\rho$ can be converted to 
the dust production rate in kg s$^{-1}$, $Q_{Af\rho}$, 
using \cite{Cremonese2020ApJ} \citep[see also,][]{Fulle2004come.book}, 
\begin{equation}
Q_{Af\rho} \approx (A f \rho)~\frac{2\,a_{\rm d}\,\rho_{\rm d}\,v_{\rm d}}{3\,p_{\rm v}}.  
\label{Cremonese2020Qdust}
\end{equation} 
We adopt the same values of 
$a_{\rm d}$, $\rho_{\rm d}$, $v_{\rm d}$ and $p_{\rm v}$ as above.   
The results of the $Af\rho$-method ($Q_{Af\rho}$) are also shown in Table~\ref{final_results}. 
We obtain $Q_{\rm Af\rho}$ = 1.0$\pm$0.2$\times$10$^{-2}$\,kg\,s$^{-1}$ in Visit~$A$ ($R_{\rm h}$ = 1.20\,au) 
and $Q_{\rm Af\rho}$ = 4.0$\pm$0.2$\times$10$^{-2}$ in Visit~$B$ ($R_{\rm h}$ = 1.01\,au), respectively.

$Q_{\rm dust}$ and $Q_{Af\rho}$ show consistency in Visit~$A$,  
while in Visit~$B$ they show a difference by a factor of two.  
This difference in Visit~$B$ could be caused due to a few reasons.
One is by inapplicability of the Halley-Marcus phase function ($\Phi_{\rm HM}$) 
at larger phase angles.   
Visit~$A$ had a phase angle $\alpha \sim$ 50$^{\circ}$, 
while in Visit~$B$ it had a larger $\alpha \sim$ 70$^{\circ}$. 
The Halley-Marcus is a composite function,  
having a critical point at $\alpha$= 55$^{\circ}$.
It combines the use of the Halley-function at smaller 
phase angles \citep{Schleicher1998Icar} 
with the Marcus-function at larger angles \citep{Marcus2007ICQ_29_39M}\footnote{https://asteroid.lowell.edu/comet/dustphase/details}\citep[see also,][]{Marschall2022AA}. 
This might influence the $Af\rho$-method for deriving dust production rate 
as seen in Equations~(\ref{Afro}) and (\ref{Cremonese2020Qdust}).  
Another cause could be the phase angle dependence 
of the dust velocities.  
The simulation conducted for JFC 67P/Churyumov-Gerasimenko (when at $R_{\rm h}$ = 1.24\,au) 
finds that the speed of $\micron$-sized dust particles 
at $\alpha \sim$70$^{\circ}$ reaches only $\sim$80\% 
of those at $\alpha \sim$ 50$^{\circ}$ \citep[Figure~8, right panel in][]{Agarwal2023arXiv}.  
This slower speed at higher phase angle could influence 
the both methods, potentially overestimating the current production rates. 
For Visit~$B$, we attempted to adopt $\sim$80\% of $v_{\rm d}$ in Visit~$A$, 
which is $v_{\rm d}$ = 9.2\,m\,s$^{-1}$, finding 
$Q_{\rm dust}$ = 2.0 $\pm$ 0.1 $\times$10$^{-2}$\,kg\,s$^{-1}$ and 
$Q_{Af\rho}$ = 3.0 $\pm$ 0.1 $\times$10$^{-2}$\,kg\,s$^{-1}$, respectively.    
The difference is mitigated by the $25\%$ reduced value of $Q_{Af\rho}$.  
This is a preliminary analysis, implying the need for more precise calculations.  
However, it finds that the cross-section method 
is less sensitive to phase angle variations in this case. 
The comparison between $Q_{\rm dust}$ and $Q_{Af\rho}$ 
confirms that both methods can yield 
similar estimates, typically within a factor of a few.

The dust-to-gas production ratio, $f_{\rm dg}$, is derived using \citep{Jewitt2014AJ},  
\begin{equation}
f_{\rm dg} = \frac{Q_{\rm dust}}{\pi r_{\rm s}^2 F_{\rm s}}, 
\label{f_dg}
\end{equation}
where 
$r_{\rm s}$ is the effective radius of the sublimating ice patch area, 
$F_{\rm s}$ (kg\,m$^{-2}$\,s$^{-1}$) is the calculated specific mass-loss rate, 
and $Q_{\rm dust}$ (kg s$^{-1}$) is the measured dust production rate obtained above.  
Substituting the values derived in the previous sections 
for $r_{\rm s}$, $F_{\rm s}$ and $Q_{\rm dust}$ 
yields 2\,$\leqslant\,f_{\rm dg}\,\leqslant$\,6 
(approximated from 1.9\,$\pm$\,0.2\,$\leqslant f_{\rm dg} \leqslant$\,5.6\,$\pm$\,1.0).  
This result is comparable with measurements from 67P, where Rosetta found 
$f_{\rm dg}$ = 4\,$\pm$\,2 or $\sim$6 \citep{Rotundi2015_67P, Fulle2016ApJ}.  
2P/Encke, 
the parent body of the Taurid meteoroid stream,  
has 10 $\leqslant$  $f_{\rm dg}$  $\leqslant$ 30 \citep{Reach2000Icar}.

\begin{figure*}[ht!]
\epsscale{1} \plotone{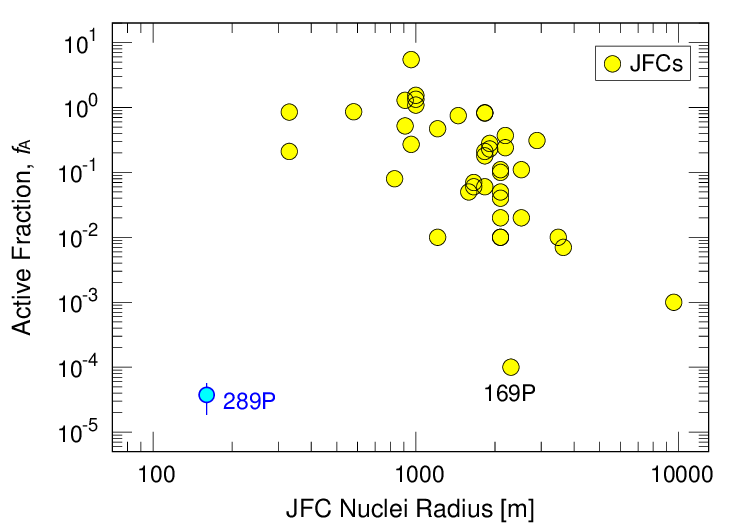} 
\caption{
Relationship between JFC nuclei radius 
and fractional active area, $f_{\rm A}$.  
The $f_{\rm A}$ of 289P is estimated near its perihelion, 
despite being among the lowest in JFCs, 
comparable to the inactive state of 169P.     
}
\label{f_A}
\end{figure*}

The fraction of active area (ice patch) 
on the nucleus surface, $f_{\rm A}$, is 
derived using \cite{LuuJewitt1992Icar}, 
\begin{equation}
f_{\rm A} = \frac{Q_{\rm dust}}{4 \pi r_{\rm n}^2\,f_{\rm dg}\, F_{\rm s}}.  
\label{fraction}
\end{equation}
Using the above parameters, 
we find $f_{\rm A}$ $\sim$ 5.1$\times10^{-5}$ in Visit~$A$ and 
$f_{\rm A}$ $\sim$ 2.4$\times10^{-5}$ in Visit~$B$, respectively.  
Figure~\ref{f_A} shows the radius (m) versus fractional active area 
$f_{\rm A}$ for 289P, along with determinations of $f_{\rm A}$ 
for other JFCs \citep{Tancredi2006Icar,Kasuga_et_al2010}.  
The small active surface fraction of 289P is obvious, 
with the averaged value $f_{\rm A}$ = 3.8\,$\pm$\,1.9\,$\times\,10^{-5}$, 
compared to the km-sized JFCs.   
The similar small fraction ($f_{\rm A} < 10^{-4}$) 
is found in 169P/NEAT, 
which remained in its dormant state 
even at $R_{\rm h}$\,=\,1.4\,au \citep{Kasuga_et_al2010}.   
Small $f_{\rm A}$ is also found in 
28P/Neujmin ($f_{\rm A}$ = 0.001), 
likely caused by dust mantle smothering 
the active area due to the gravity of 
its large 10\,km radius body \citep{Tancredi2006Icar}.
The values $f_{\rm A} >$\,1 are seen in 
21P/Giacobini-Zinner, 24P/Schaumasse and 
73P/Schwassmann-Wachmann, indicating
ice sublimation both from the nucleus and from icy grains 
ejected from nucleus \citep{Jewitt2022AJ_LPC}.  
The earlier study suggests that 
as nucleus radius decreases, 
$f_{\rm A}$ approaches unity
on sub-km short-period comets, 
but selection bias possibly remains \citep{Jewitt2021AJ}.  
Obtaining more samples for 100\,m-sized bodies 
would be helpful while, 289P currently exhibits 
the smallest $f_{\rm A}$ among the reported JFCs.

\subsection{Gas (CO$_2$ and CO) Production Rate: $W2$}
\label{W2Q}

The $W2$ (4.6$\micron$)-band is utilized for 
investigating the volatile compositions of active comets.   
Excess flux density in this band is interpreted 
as arising from the emission of gas molecules, primarily     
the CO$_2$ $\nu_3$ vibrational fundamental band (4.26$\micron$) 
and the CO $v(1-0)$ rovibrational fundamental bands (4.67$\micron$).  
These volatile species have low sublimation temperatures of 20$-$100\,K 
and are preserved in frozen ices or trapped as gases within 
the cometary nuclei, preferably 
at distant place beyond $R_{\rm h} >$\,3\,au \citep{Prialnik2004come.book,Womack2017,BouzianiJewitt2022}.  
Space-based imaging studies have placed the upper limits on 
those of production rates in active comets with $R_{\rm h}$ $\lesssim$\,12\,au \citep{Pittichova2008AJ,Bauer2011ApJ,Bauer2015ApJ,Stevenson2015ApJL,Bauer2021PSJ,Rosser2018AJ,Gicquel2023PSJ,Milewski2024AJ}.  
The individual gas emission bands, 
particularly those of CO$_2$, have been detected in comets
using satellite spectroscopy \citep{Ootsubo2012ApJ}.

However, we find no excess signal in 289P after removing 
reflected sunlight and thermal emission (Section~\ref{model}). 
This indicates that CO$_2$ and CO gas production 
is negligibly small or absent in 289P. 
The small nucleus size ($r_{\rm n}$\,$\sim$\,160\,m), 
the small $q$\,=\,0.96\,au, 
and 
the short $P_{\rm orb}$\,=\,5.3\,yr 
likely drive its rapid sublimation and loss of CO$_2$ and CO.  
The core temperature around the 289P orbit 
\citep[Equation~(4) of][]{JewittHsieh2006} 
is $T_{\rm core}$\,$\sim$\,180\,K.  
Even assuming a highly porous nature for 289P,  similar to 67P, 
and a low thermal diffusivity 
$\kappa$\,$\sim$\,10$^{-8}$$-$10$^{-7}$~m$^2$~s$^{-1}$ (from Rosetta, Appendix~\ref{kappa}), 
the longest timescale for heat diffusion into the core 
is $\tau_{\rm c}$~$\sim$~8\,$\times$\,10$^{3-4}$\,yr.  
This suggests that solar heat can reach into 
the body core much sooner than the end of 
the dynamical lifetime of short-period comets $\sim$5\,$\times$\,10$^{5}$\,yr \citep[][]{Duncan2004}.  
Consequently, CO$_2$ and CO are unlikely to be 
preserved, and mostly lost within the 289P nucleus. 

Note that water-ice (H$_2$O) can still be contained in comet 289P.  
While the $T_{\rm core}$ exceeds the calculated sublimation 
temperature of pure water-ice 150\,K \citep{Yamamoto1985}, 
it is comparable to or lower than 180\,K~$\sim$~210\,K 
range reported for dust-mixed water-ice ($f_{\rm dg}$\,=\,3) 
based on surface spectroscopic data of JFC 103P/Hartley~2 \citep[EPOXI,][]{Yue2023JGRE}.   
This suggests that dust-mixed water ice within 289P (2 $\leqslant f_{\rm dg} \leqslant$ 6)
would be more stable than pure ice, 
potentially hindering rapid sublimation of water ice.    
The observed sublimation-driven activity in 289P 
is weak and looks to be nearly end, while there remains a chance 
of its longevity, provided that its steady state is retained.     
Overall, comet 289P is unlikely to be a reservoir
for CO$_2$ and CO, but rather for water-ice.

\section{Discussion}
 
\subsection{The Phoenicid Stream Mass}
\label{stream}

We discuss three types of mass supply to 
form the Phoenicid meteoroid stream: 
steady, destructive and outburst.  
The sources considered are from 
comet 289P and/or its proposed 
precursor body.

\subsubsection{Steady Mass Supply}
\label{steadysupply}

The estimated mass-loss rates in dust (dust production rates) from 
observations of 289P can be compared with the mass of the Phoenicid stream.  
\cite{{JenniskensLyytinen2005}} estimated the stream 
mass $\sim 10^{11}$\,kg from the 1956 Phoenicids, claiming 
that this could be consistent with the mass-loss activity (brightness) 
of an unconfirmed, few-km-scale precursor comet 
during a single orbital passage.  
The actual mass-loss rates are 
0.01$-$0.02\,kg\,s$^{-1}$ (Section~{\ref{W1Q}}), 
two to four orders of magnitude smaller than 
those of active JFCs \citep[e.g.][]{Gillan2024PSJ}.   
This discrepancy suggests an alternative approach 
to estimating the Phoenicid stream mass: using  
the parent body mass, $M_{\rm n}$, 
as a starting point.    
The other meteoroid streams 
show an approximate order unity agreement 
between the stream mass and the parent body mass.
For example, the Taurid stream associated with comet 
2P/Encke is estimated to have a mass of $\sim$10$^{14}$\,kg.  
Similar mass relationships are found in the 
$\alpha$-Capricornid stream (from comet 169P/NEAT) 
and the Quadrantid stream (from 2003\,EH$_1$), 
with estimated masses of $\sim$10$^{13}$\,kg each 
\citep[][]{KasugaJewitt2019}.  
Given this consistency, we propose 
using the nucleus mass of 289P for 
the Phoenicid stream mass.  
The estimated nucleus mass of 289P 
is $M_{\rm n} \sim 2 \times 10^{10}$\,kg 
(with factors of several uncertainty) \citep{Jewitt2006}, 
which is approximately one-fifth of the calculated stream mass $\sim 10^{11}$\,kg \citep{{JenniskensLyytinen2005}}. 
By scaling $M_{\rm n}$ up and down by a factor of five, 
we estimate the possible range for the Phoenicid stream mass: $4\times10^{9}$ $\lesssim  M_{\rm s} \lesssim 10^{11}$\,kg.  

The Phoenicid stream is estimated to have been formed 
within $\tau_{\rm s} \sim$\,300\,yrs \citep[][]{Watanabe2005PASJ}. 
Assuming the observed mass-loss rate of 0.02\,kg\,s$^{-1}$, 
the total stream mass would be only $\sim 1.9 \times 10^{8}$\,kg, about 
an order of magnitude smaller than the possible stream mass range ($10^{9} \sim10^{11}$\,kg).   
Therefore, we conclude that the Phoenicid stream was not solely 
produced through steady mass-loss at the observed rates.  
 
\begin{figure*}[htbp]
\epsscale{1} \plotone{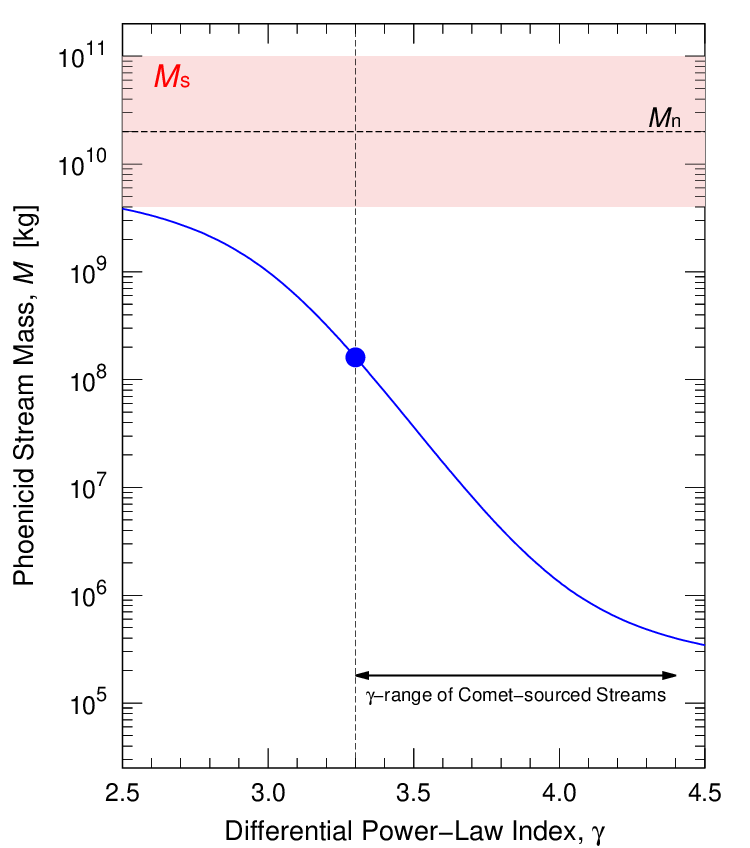} 
\caption{Phoenicid stream mass is plotted 
as a function of the differential power-law index, 
$\gamma$, as a blue solid line ($\gamma$~$\ne$~3.0, 4.0 in Equation~(\ref{streammass})).   
The possible stream mass range ($M_{\rm s}$),  
based on the nucleus mass \citep[$M_{\rm n}$,][]{Jewitt2006}
and 
the estimated stream mass \citep[$\sim$\,10$^{11}$\,kg,][]{JenniskensLyytinen2005}, 
is shown as a reddish horizontal band.  
The Phoenicid stream mass 
is determined at $\gamma$ = 3.3 (blue circle), 
which lies closest to the possible stream mass range,  
within the plausible index range 3.3\,$\leqslant \gamma \leqslant$\,4.4 (double-headed arrow) 
derived from the observed comet-sourced meteoroids \citep[][]{Blaauw2011MNRAS,Egal2022MNRASobs,Egal2022MNRAS,Koten2023AA,Moorhead2024}.  
}
\label{Stream_mass_fig}
\end{figure*}

Next, we focus on the size of the dust particles (meteoroids) 
and their distribution in the Phoenicid stream.    
In general, cometary streams mostly consist of 
near mm$\sim$cm-sized meteoroids (compact aggregates), 
as revealed by infrared space 
observations \citep[2P, 73P, etc,][]{SykesWalker92,Reach2000Icar,Reach07,Reach2009Icar}.
Comparable size ranges, from nearly mm up to 10\,cm, 
have been measured from the observations of meteor showers, 
such as the Taurids from 2P \citep{Egal2022MNRASobs} 
and the 2022 $\tau$-Herculids (young stream, $\sim$\,30\,yrs old) from 73P \citep[][]{Koten2023AA}.  
These larger particles, potentially containing much more mass, 
play a major role in determining the total stream mass.
These observations imply that the Phoenicid stream is likely 
composed of meteoroids with a similar size distribution, 
ranging from nearly mm to 10\,cm.  
We assume a distribution of dust particle radii 
follows a differential power law index $\gamma$, such 
that number of particles having radius between $a$ 
and $a$+$da$ is written as 
$n(a)da = \Gamma\,a^{-\gamma}da$, with a constant $\Gamma$.  
The integrated mass of the particles, $M$, between radii $a_{1}$ 
and $a_{2}$ over the last centuries is expressed as, 
\begin{equation}
M = \frac{4}{3}\,N_{\rm orb}\,\rho_{d}\,C_{\rm d}\,\bar{a}, 
\label{streammass}
\end{equation}
where $N_{\rm orb}$ is number of perihelion passages of 289P, 
$C_{\rm d}$ is the cross-section of all the dust particles.  
The $\bar{a}$ is the area-weighted mean particle radius, 
calculated by \citep{Jewitt2023AJ}, 
\begin{equation}
\bar{a} = \frac{\int_{a_1}^{a_2} a^{3-\gamma} da}{\int_{a_1}^{a_2} a^{2-\gamma} da}, 
\label{a_bar}
\end{equation}
where 
$a_1$ is the minimum particle radius,   
$a_2$ is the maximum particle radius, 
and $\Gamma$ is eliminated.  
We set 
$N_{\rm orb}$\,$\sim$\,57 ($\tau_{\rm s}$/$P_{\rm orb}$ = 300\,yrs / 5.3\,yrs), 
$C_{\rm d}$ \,=\,1.5\,km$^2$ (Visit~$A$ in Table~\ref{final_results}), 
and 
$a_1$ = 1\,$\micron$ ($W1$-like),    
and 
$a_2$ = 10\,cm (fireball-class meteoroids and the SSA model in Figure~\ref{SSA_fig}).  
With these values, we plot Equation~(\ref{streammass}) as 
a function of $\gamma$ in the range 2.5 $\leqslant \gamma \leqslant$ 4.5, 
as shown in Figure~\ref{Stream_mass_fig}.  
As the power law index ($\gamma$) decreases, 
the stream mass increases and approaches to 
the possible mass range due to 
the shift in the particle size distribution towards 
larger particles.  
A relevant comparison can be made with the 
size distribution of the observed meteoroid streams.   
The index range, 
3.3\,$\leqslant \gamma \leqslant$\,4.4, is estimated 
from the 1956 Phoenicids \citep[$\gamma$\,$=$\,4 from][]{Venter1957,Moorhead2024} 
and 
the other meteor showers associated 
with cometary sources \citep[][]{Blaauw2011MNRAS,Egal2022MNRASobs,Egal2022MNRAS,Koten2023AA}.   
We apply these results to estimate a plausible 
Phoenicid stream mass $M \sim$1.6$\times$10$^{8}$\,kg 
at $\gamma$ = 3.3, still an order 
of magnitude lower than the possible range (Figure~\ref{Stream_mass_fig}).  
The difference in mass ($\sim$3.84$\times$10$^{9}$\,kg) 
corresponds to 
a spherical fragment having a radius $\sim$\,100\,m 
with an assumed density of 1000\,kg\,m$^{-3}$ (=\,$\rho_{\rm n}$), 
equivalent to a 10\,m-thick surface shell on the 160\,m-radius nucleus of 289P.    
The leading conclusion is again that steady mass-loss 
near perihelion, even considering larger meteoroids and the distribution, 
cannot produce the Phoenicid stream.

\subsubsection{Destructive Mass Supply}

\begin{figure*}[htbp]
\epsscale{1} \plotone{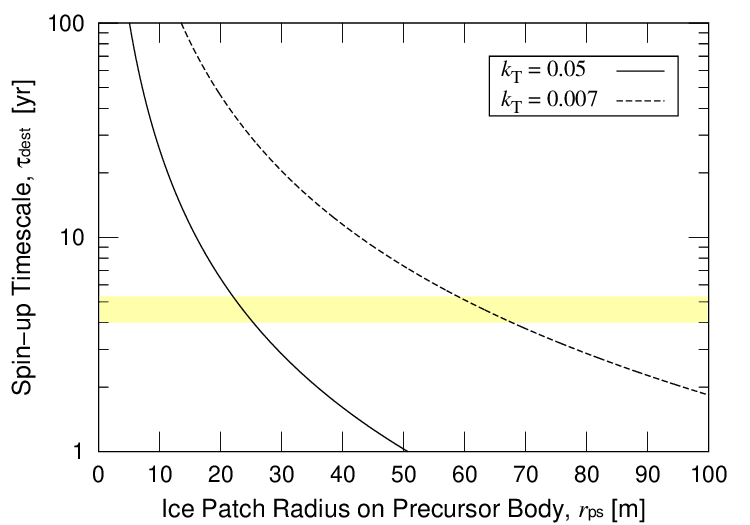} 
\caption{
Precursor spin-up timescale 
is plotted as a function of 
radius of ice sublimating area (patch), 
$r_{\rm ps}$, from Equation~(\ref{tau_dest}).  
Solid and dashed curves (black) show  
$k_{\rm T}$\,=\,0.05 and 0.007 \citep{Jewitt1997EMP,Jewitt2021AJTorques}, respectively.  
A horizontal band (yellow) indicates 
the range of spin-up timescales constrained 
both by orbital period and empirical observations.    
The two curves within the band highlight 
the parameter space where 
rotational destruction can occur.  
}
\label{tau_dest_fig}
\end{figure*}

The Phoenicid meteoroid stream is likely replenished 
by the breakup of the 289P precursor body, 
but the exact cause remains unknown.    
Tidal forces from Jupiter at aphelion 
were unlikely 
due to their large distance (0.07$\sim$0.26\,au) 
during their 1626$-$1819 interaction \citep{JenniskensLyytinen2005}. 
Other mechanisms must have led to 
the breakup. 
Here, we focus on the small size of 
the precursor body with 
the presumable radius $r_{\rm p}$\,$\sim$\,170\,m (Section~\ref{steadysupply}).     
Sub-kilometer active nuclei are expected to rapidly 
disintegrate due to rotational instability 
induced by outgassing torques, 
on timescales comparable to their orbital periods 
\citep{Samarasinha2013,Jewitt2021AJTorques}.   
We examine this possibility and its past active state.   
Outgassing may have spun-up the precursor nucleus, 
leading to its destruction on a timescale \citep{Jewitt1997EMP,Jewitt2021AJTorques},  
\begin{equation}
\tau_{\rm dest} = \left(\frac{16\pi^2}{15}\right)\,\left(\frac{\rho_{\rm n}\,r_{\rm p}^4}{k_{\rm T}\,v_{\rm g}\,P_{\rm rot}^\prime}\right)\left(\frac{1}{\overline{F_{\rm s}}\,\pi\,r_{\rm ps}^2}\right).  
\label{tau_dest}
\end{equation}
In this equation, 
$k_{\rm T}$ is the dimensionless effective moment arm of the outgassing
\footnote{ 
Quantity $k_{\rm T}$\,=\,1 corresponds to 
tangential emission, while $k_{\rm T}$\,=\,0 
represents perfectly central outgassing \citep{Jewitt1997EMP}.  
},  
$P_{\rm rot}^\prime$ is the rotational period of the precursor body, 
$\overline{F_{\rm s}}$ is the average sublimation mass-loss rate of 
water ice per unit area around its orbit,  
$r_{\rm ps}$ is the radius of the ice sublimating area (patch) on the precursor body, 
and the other parameters are as described above.  
The spin-up timescale range is set 
between 
the orbital period $P_{\rm orb} = 5.3$\,yr (Table~\ref{289P_orbit}) 
and 
the empirically estimated timescale $\sim$\,4\,yr, 
given by $\sim$100\,$r_{\rm p(km)}^2$\,yr ($r_{\rm p(km)} \sim$\,0.2 expressed in km),  
for short-period comets \citep{Jewitt2021AJTorques}.  
Rotational destruction occurs within this range ($\sim$\,4\,$-$\,5.3\,\,yr), 
primarily determined by $r_{\rm ps}$.   
Values of $k_{\rm T}$ derived from 
both models and observations are within the range 
2$\times$10$^{-4}$\,$\lesssim$\,$k_{\rm T}$\,$\lesssim$\,5$\times$10$^{-2}$ 
\citep{Jewitt1997EMP,Belton2011Icar,Drahus2011ApJ,Jewitt2021AJTorques}.  
Small nuclei with low $f_{\rm A}$, like 289P, are considered 
remnants (mostly unobserved) of destructive events 
caused by their large $k_{\rm T}$ \citep[][]{Jewitt2021AJTorques}.
Localized sublimation, as modeled for 289P (Section~\ref{patch}), 
also implies a larger $k_{\rm T}$.  
We selected two representative values: 
$k_{\rm T}$\,=\,0.05 \citep{Jewitt1997EMP} 
and the median value of $k_{\rm T}$\,=\,0.007 
based on the short-period 
active comet measurements \citep{Jewitt2021AJTorques}.   
The larger value, $k_{\rm T}$\,=\,0.05, was prioritized.  
We computed the orbitally averaged  
subsolar sublimation $\overline{F_{\rm s}}$ 
for the precursor body 
with $a$\,=\,3.05\,au and $e$\,=\,0.69 (Table~\ref{289P_orbit}), 
finding $\overline{F_{\rm s}}$\,=\,4\,$\times$\,10$^{-5}$\,kg\,m$^{-2}$\,s$^{-1}$ 
\citep[Equation~(A4),][]{Jewitt2021AJTorques}. 
To first order, we adopted 
$v_{\rm g} \sim$\,480\,m\,s$^{-1}$ near $R_{\rm h}$ = 1\,au (Section~\ref{patch}) 
evidenced by the weak distance-dependence 
of $v_{\rm g}$~$\propto$~$R_{\rm h}^{-1/4}$ \citep[][]{Biver2002EMP,Jewitt2021AJTorques}. 
Assuming $P_{\rm rot}^\prime$\,=\,$P_{\rm rot}$\,$\sim$\,8.8536\,hr (Section~\ref{P_rotation})
and 
using the above parameters, 
we plot Equation~(\ref{tau_dest}) 
as a function of $r_{\rm ps}$ in Figure~\ref{tau_dest_fig}. 
Rotational destruction would have occurred 
for $k_{\rm T}$\,=\,0.05 when 22\,m\,$\lesssim$\,$r_{\rm ps}$\,$\lesssim$\,25\,m 
and 
for $k_{\rm T}$\,=\,0.007 when 59\,m\,$\lesssim$\,$r_{\rm ps}$\,$\lesssim$\,68\,m, respectively. 
By comparison, setting perihelion distance $q$ = 0.96\,au 
finds the mass-loss rates $\sim$\,0.8\,$-$\,1.0 kg s$^{-1}$ ($k_{\rm T}$\,=\,0.05) 
or $\sim$\,6\,$-$\,8 kg s$^{-1}$ ($k_{\rm T}$\,=\,0.007), respectively.  
The corresponding fractional active area is 
$\sim$\,(4\,$-$\,5)\,$\times$\,10$^{-3}$ ($k_{\rm T}$\,=\,0.05) 
and $\sim$\,(3\,$-$\,4)\,$\times$\,10$^{-2}$ ($k_{\rm T}$\,=\,0.007), respectively. 
Both results are more than an order of magnitude 
larger than those of 289P, while comparable to typical JFCs.  
For the $k_{\rm T}$\,=\,0.05 case, 
the radius of sublimation area is at least 10\,$\times$ larger than 
$r_{\rm s}$\,$\approx$\,2\,m on the 289P nucleus (Section~\ref{patch}), 
but the low mass-loss rates are remarkable.  
The precursor body, 
despite low activity, 
could have spun-up and disintegrated rapidly.  

The destruction would have contributed to the Phoenicid stream mass.  
The exact timing of the precursor's rapid spin-up 
and subsequent destruction remains unknown, 
but we might be able to narrow it down to specific years. 
The 289P's orbit has been changed 
due to strong interactions with the Earth 
at perihelion and Jupiter at aphelion 
over short timescales \citep{Jewitt2006}.  
The dynamical modeling proves that 
dust trails ejected before 1743 are too sparse 
to explain the 1956 Phoenicids, 
while those ejected in 1760$-$1808 are the major sources \citep{Watanabe2005PASJ}.  
In 1819, the comet was already very faint \citep{Kronk2003comebook},  
suggesting that the precursor body had been disintegrated before that year.    
Given these constrains, 
the destructive event occurred probably in 1743$-$1819.     

To verify it, we examine the potential contribution 
from the precursor's sublimation-driven mass-loss during the given timeframe.  
Using $\overline{F_{\rm s}}$\,=\,4\,$\times$\,10$^{-5}$\,kg\,m$^{-2}$\,s$^{-1}$ 
and 
the largest $r_{\rm ps}$\,$\sim$\,25\, and \,68\,m (when $k_{\rm T}$\,=\,0.05 and 0.007, respectively), 
the maximum delivered mass over 76 years (=\,1819\,$-$\,1743) 
would be (0.2\,$-$\,1)\,$\times$\,10$^{9}$\,kg.  
This accounts for only 25\,\% or less of 
the minimum possible stream mass of $4\times10^{9}$\,kg ($\lesssim M_{\rm s}$).   
The $k_{\rm T}$\,=\,0.05 case would supply 
more than an order of magnitude smaller mass.       
These estimates can be consistent with 
the lack of observations before 1819 \citep{Kronk2003comebook}, 
most likely owing to its faintness due to low activity.   
Our finings strongly suggest that sublimation-driven activity 
from the precursor was 
insufficient to account for the Phoenicid stream mass.  
Thus, destructive mass supply was essential for its formation.    

The Phoenicid meteoroid stream would 
comprise dust particles produced from both 
sublimation and disintegration processes.  
Each process ejects particles 
with speeds, 
$\gtrsim$\,0.5\,m\,s$^{-1}$ from sublimation \citep[][]{Watanabe2005PASJ} 
and 
$\sim$\,0.2\,m\,s$^{-1}$ ($\approx$ escape velocity) from disintegration, 
differing by a factor of few at least.  
Both particle types may have undergone 
similar dynamical evolution but may not have 
reached the Earth's orbit simultaneously, 
depending on the timing of the destructive event.  
\cite{{JenniskensLyytinen2005}} 
estimated the stream mass ($\sim$\,$10^{11}$\,kg) 
based on the 1819 trail and particle sizes $\sim$\,1\,mm.   
However, during the 1956 event, 
the 1819 trail was more than twice farther away 
from the Earth orbit ($\delta d_{\rm E}$ = 1.4\,$\times$\,10$^{-3}$\,au)  
compared to the 1760$-$1808 trails ($\delta d_{\rm E}$ = 6.3\,$\pm$\,0.8$\times$\,10$^{-4}$\,au), 
suggesting its limited contribution possibly only to 
the shower's sub-peak \citep[Table~1 and Figure~1,][]{Watanabe2005PASJ}.  
To estimate the stream mass, they assumed 
a broader dust trail along the Earth's path, 
which is the cross-sectional area about 5 $\times$ larger 
than that predicted from the distance to the 1760$-$1808 trails ($\delta d_{\rm E}$), 
and integrated it along the comet's orbit.  
This approach would have potentially 
encompassed both types of particles, 
regardless of their minor orbital differences.

\subsubsection{Outburst Mass Supply}

In 2013, 289P showed the distant activity with its coma 
and a tail at $R_{\rm h}$ = 3.88\,au \citep{Williams2013CBET}, 
which produced the mass of 
$M_{\rm d}^\prime$ = 3.1~$\times$~10$^{5}$\,kg (Appendix~\ref{2013Activity}).  
The event was likely an unexpected outburst of mass production, 
corresponding to $<$\,10$^{-4}$ of the total stream mass.  
This is still too small to significantly contribute to the total stream mass, 
although there might be unseen similar events more.  

All considerations here, 
the formation of the Phoenicid stream is 
likely a result of multiple mass-loss processes 
from comet 289P, including 
steady-state mass-loss near perihelion, 
unexpected outbursts, and, most importantly, 
the disintegration of the precursor body.

\subsection{Nongravitational Acceleration}
\label{nga}

Nongravitational acceleration (NGA) on comets can 
serve as an indicator for estimating outgassing levels (mass-loss in gas).  
The NGA is decomposed into three types of 
independent parameters, 
$A_1$ in radial direction, 
$A_2$ in transverse (tangential component of the Yarkovsky force, 
influencing the object's movement along its orbital path based on its rotation), 
$A_3$ in perpendicular to the plane of 
the orbit \citep{Marsden1973AJ}.  
These parameters are calculated 
only when necessary to reconcile discrepancies 
between gravitational models and astrometric data, 
otherwise set to zero \citep{Jewitt2022AJ_LPC}.
The radial component ($A_1$) is generally 
the largest due to the concentration of 
cometary outgassing on the heated, 
sunlit side of the nucleus. 
The resulting recoil force, 
primarily driven by the sublimation of water ice, 
acts directly away from the Sun.  

In 2019 October$-$November,  
289P was returning to its perihelion ($R_{\rm h}$\,=\,1.45\,$\rightarrow$\,1.05\,au), 
and the NGA parameters were obtained 
as $A_1$ = $-$9.0$\times$10$^{-10}$\,au\,day$^{-2}$ 
and $A_2$ = $-$8.2$\times$10$^{-11}$\,au\,day$^{-2}$.\footnote{MPEC~2019-W179.~\url{https://www.minorplanetcenter.net/mpec/K19/K19WH9.html}.}
The negative $A_1$-value suggests 
inward acceleration towards the Sun during its inbound, 
likely caused by a change in angular velocity driven by 
outgassing increase \citep{Yeomans2004comebook,Jewitt2021AJTorques}.    
The absolute value is large enough to strongly 
support its cometary outgassing activity.  
If 289P were in the absence of sublimation of water ice,  
resembling a bare asteroid, 
the radial component should be 
$A_1$\,=\,1.0$\times$10$^{-12}$\,au\,day$^{-2}$ 
at $R_{\rm h}$\,=\,1.05\,au \cite[Equation~(2) of][]{Hui_Jewitt2022AJ}.   
This is three orders of magnitude smaller than 
the determined value from the astrometry.   
Therefore, 289P undoubtedly undergoes outgassing, 
consistent with the presence of sublimating ice.  

\begin{figure*}[ht!]
\epsscale{1} \plotone{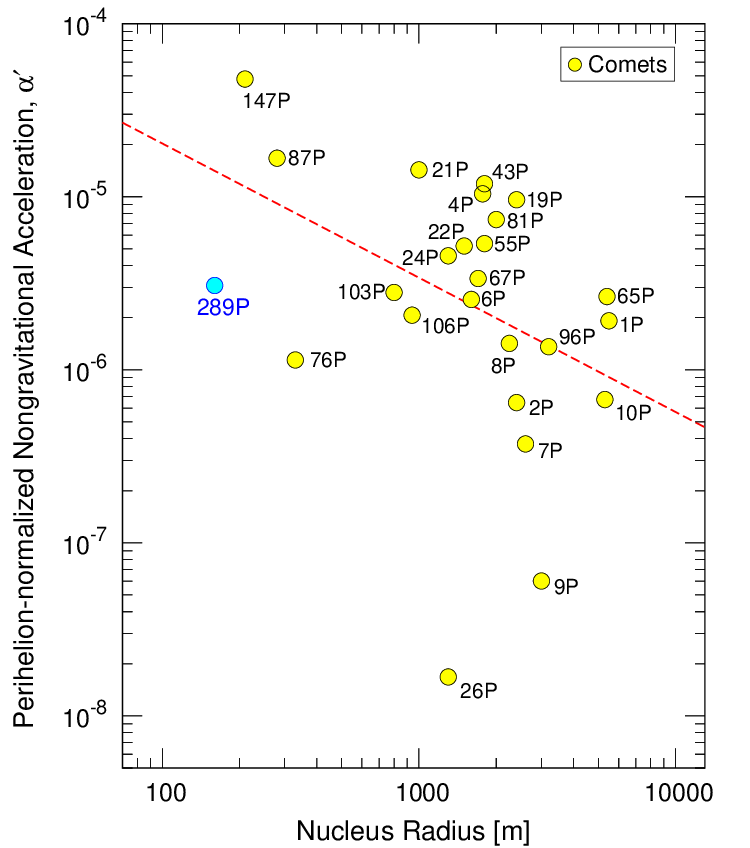} 
\caption{
Nucleus radius versus perihelion-normalized 
nongravitational acceleration of 289P (blue symbol) 
and comparison with those of well-studied 
short period comets (yellow symbols).   
The dashed red line indicates the trend found by fitting 
from the latter data,   
which is 
Halley-type (1P and 55P), 
Encke-type (2P, 87P, and 147P), 
and
JFCs (4P, 6P, 7P, 8P, 9P, 10P, 19P, 21P, 22P, 24P, 26P, 43P, 65P, 67P, 76P, 81P, 96P, 103P, and 106P).  
}
\label{alpha_p}
\end{figure*}

The nongravitational acceleration, $\alpha_{\rm NG}$, 
is given as a function of sublimation rate of water ice, $g$ ($R_{\rm h}$), 
and expressed by \citep{Marsden1973AJ} 
\begin{equation} 
\alpha_{\rm NG} = g(R_{\rm h})(A_1^2 + A_2^2 + A_3^2)^{\frac{1}{2}}. 
\label{ng}
\end{equation} 
The function $g(R_{\rm h})$ is defined by 
\begin{equation}
g(R_{\rm h}) = \alpha_{\rm M} \left( \frac{R_{\rm h}}{R_{\rm 0}} \right)^{-m}\,\left[\,1+ \left(\frac{R_{\rm h}}{R_{\rm 0}} \right)^{n}\,\right]^{-k}, 
\label{g_func}
\end{equation}
where 
$R_{\rm 0}$ = 2.808\,au, 
$m$ = 2.15, 
$n$ = 5.093, 
$k$ = 4.6142, 
and 
$\alpha_{\rm M}$ = 0.1113
are constants determined from a fit to a model of sublimation of water ice.  
Substituting $R_{\rm h}$=1.01\,au (Visit~$B$),
$A_1$ = $-$9.0$\times$10$^{-10}$\,au\,day$^{-2}$ , $A_2$ = $-$8.2$\times$10$^{-11}$\,au\,day$^{-2}$, 
and $A_3$ = 0\,au\,day$^{-2}$ into Equations~(\ref{ng}) and (\ref{g_func}), 
we find $\alpha_{\rm NG}$ = 1.8$\times$10$^{-8}$\,m\,s$^{-2}$ on 289P, 
about 0.0003 \% of the solar gravitational acceleration ($g_{\odot}$).  
The result from 289P can be compared 
with those of well-characterized 
short-period comets.  
We focus on the perihelia, where 
the strongest outgassing  
produces the maximum 
nongravitational acceleration.   
The nucleus radii and the determined 
parameters $A_1$, $A_2$ and $A_3$ 
are taken from the NASA/JPL Small-Body Database Lookup.\footnote{\url{https://ssd.jpl.nasa.gov/tools/}}
The $\alpha_{\rm NG}$ is normalized by 
the solar gravitational acceleration 
at each comet's perihelion distance (i.e. perihelion-normalized NGA), 
as given by \citep{Jewitt2021AJ}, 
\begin{equation} 
\alpha_{\rm NG}^\prime = \frac{g(q) q^2}{GM_\odot} (A_1^2 + A_2^2 + A_3^2)^{\frac{1}{2}},  
\label{ng_prime}
\end{equation}
where $G$ is the gravitational constant, $M_\odot$ is the Solar mass, 
and $q$ is the perihelion distance in meters.   
Substituting $q$ = 0.96\,au into Equation~(\ref{g_func}), we obtain $g(q)$\,=\,1.097.  
Using Equation~(\ref{ng_prime}) with the parameters above, we find $\alpha_{\rm NG}^\prime$\,=\,3.1$\times$10$^{-6}$.  
Likewise we derive $\alpha_{\rm NG}^\prime$ for known comets: two Halley-type (1P and 55P), 
three Encke-type (2P, 87P, and 147P), 
and 
nineteen JFCs (4P, 6P, 7P, 8P, 9P, 10P, 19P, 21P, 22P, 24P, 26P, 43P, 65P, 67P, 76P, 81P, 96P, 103P, and 106P).   
Figure~\ref{alpha_p} shows the resulting relationship 
between nucleus radius and $\alpha_{\rm NG}^\prime$.  
We find Halley-type of $\alpha_{\rm NG}^\prime$\,=\,3.6\,$\pm$\,2.4$\times$10$^{-6}$, 
Encke-type of $\alpha_{\rm NG}^\prime$\,=\,2.2\,$\pm$\,2.4$\times$10$^{-5}$, 
and 
JFCs (except 289P) of $\alpha_{\rm NG}^\prime$\,=\,4.3\,$\pm$\,4.3$\times$10$^{-6}$, 
respectively.  
The median of known comets is $\alpha_{\rm NG}^\prime$\,=\,2.7$\times$10$^{-6}$.  
The 289P ($\alpha_{\rm NG}^\prime$\,=\,3.1$\times$10$^{-6}$) falls within the range 
10$^{-7}$$\lesssim$~$\alpha_{\rm NG}^\prime$~$\lesssim$10$^{-5}$,
where most km-sized comets demonstrate \citep{Jewitt2021AJ}.
In contrast, 289P shows a moderately 
lower value among the other small-sized nuclei with 
radii of 100(s)\,meters.  
To investigate the relationship between radius 
and $\alpha_{\rm NG}^\prime$, 
we conducted a least-squares fit of a power law 
to the data for known comets (yellow symbols).  
The resulting dashed line suggests a trend 
of increasing $\alpha_{\rm NG}^\prime$ with 
decreasing radius. 
It is reasonable to expect that 
smaller nuclei would experience significantly 
greater acceleration if the outgassing levels 
were comparable across all sizes.   
While this trend is evident, 289P shows 
about an order of 
magnitude smaller $\alpha_{\rm NG}^\prime$ 
than the fitted line.  
This suggests that 289P's outgassing level is 
rather lower than those of known comets 
with comparable size.   
Note that this assessment is based on 
the limited samples of sub-km nuclei (several of them).   
To establish a more statistically robust comparison,  
deriving $\alpha_{\rm NG}^\prime$ 
from a larger sample of sub-km nuclei, 
such as those from the upcoming LSST survey, 
would provide valuable insights.  
While this current estimate of the $\alpha_{\rm NG}^\prime$-trend 
might be approximate and influenced by selection bias, 
it can still serve as a useful benchmark for 
assessing outgassing levels across various comet populations.







\section{Summary} 

We present NEOWISE observations 
for comet 289P/Blanpain near perihelion 
taken on two opportunities: 
UT 2019-10-30 ($R_{\rm h}$ = 1.20\,au, inbound) 
and 
UT 2020-01-11/12 ($R_{\rm h}$ = 1.01\,au, outbound).   
The near-infrared data, 3.4$\micron$ ($W1$) 
and 4.6$\micron$ ($W2$), 
are used to analyze its faint activity driven 
by sublimation of water ice and resulting product, 
the Phoenicid meteoroid stream.  
Based on the 1956 Phoenicids, 
considering both meteor observations and trail theory,   
we set proper constraints for dust production from 
the limited sublimating ice patch area.   
The following are findings.

\begin{enumerate}
\item   The ejected dust mass is $M_{\rm d}$ = 4100\,$\pm$\,200\,kg (inbound)
           and 1700\,$\pm$\,200\,kg (outbound), respectively.  
\item   The dust production rates are 
           $Q_{\rm dust}$ = 0.01\,$-$\,0.02\,kg\,s$^{-1}$ 
           while 289P orbits from heliocentric 
           distance $R_{\rm h}$ = 1.20 to 1.01\,au around perihelion, 
           indicating weak activity as previously observed.  
\item   The dust-to-gas production ratio ranges   
           2\,$\leqslant\,f_{\rm dg}\,\leqslant$\,6.  
\item   The fractional active area $f_{\rm A}$ = 3.8$\pm$1.9$\times10^{-5}$ 
           is the smallest in the active Jupiter family comets yet reported. 
\item  The absence of 4.6$\micron$ ($W2$) excess suggests 
          that 289P contains negligible amount of CO$_2$ and CO. 
\item  The light curve, derived from the weighted mean of $W1$ and $W2$ magnitudes,  
          shows a distinct peak amplitude with a possible rotation period 
          $P_{\rm rot}$\,=\,8.8536\,$\pm$\,0.3860\,hr,   
          but definitive conclusion is pending due to the low significance level of $\sim$\,30\,\%.  
\item  The perihelion-normalized 
          nongravitational acceleration, $\alpha_{\rm NG}^\prime$ = 3.1$\times$10$^{-6}$,   
          is approximately an order of magnitude smaller than the trend 
          observed in well-studied comets, consistent with the weak outgassing.  
\item  Current dust production from 289P, despite a plausible set of 
          particle size and distribution, is an order of magnitude 
          too small to account for the probable mass of 
          the Phoenicid stream within $\sim$300\,year dynamical lifetime.       
          289P is most likely a remnant comet of a sub-km precursor body that 
          may have been a low active JFC.  
          Outgassing quickly induced rotational destruction,  
          led to additional mass supply probably in 1743$-$1819.   
\end{enumerate}


\begin{acknowledgments}
TK is grateful to Junji Nakamura for observing  
the 1956 Phoenicid meteor shower 
and sharing his firsthand story.  
The author thanks 
Jun-ichi Watanabe, Mikiya Sato, Chie Tsuchiya, 
David Jewitt, Joseph R. Masiero, Man-To Hui, Dave G. Milewski, and Budi Dermawan, 
for support.      
This publication makes use of data products from the {\it Wide-field Infrared Survey Explorer}, 
which is a joint project of the University of California, Los Angeles, and 
the Jet Propulsion Laboratory/California Institute of Technology, funded by 
the National Aeronautics and Space Administration. 
Also, this publication makes use of data products from the Near-Earth Object Wide-field 
Infrared Survey Explorer (NEOWISE), which is a joint project of the Jet Propulsion 
Laboratory/California Institute of Technology and the University of Arizona. NEOWISE 
is funded by the National Aeronautics and Space Administration.
This publication uses data obtained from the NASA Planetary Data System (PDS). 
This research has made use of data and services provided by the International Astronomical 
Union's Minor Planet Center.  
This research has made use of the NASA/IPAC Infrared Science Archive, which is 
funded by the National Aeronautics and Space Administration and operated by 
the California Institute of Technology, 
under contract with the National Aeronautics and Space Administration.
We would like to address special thanks to anonymous reviewers and to Maria Womack for scientific editor. 
\end{acknowledgments}

%

\vspace{5mm}
\facilities{NEOWISE, WISE} 





\appendix

\section{Thermal Diffusivity}
\label{kappa}

We inferred the thermal diffusivity of 289P, $\kappa$ (m$^2$ s$^{-1}$), 
from thermal inertia measurements of JFC 67P by 
Rosetta's VIRTIS (Visible InfraRed and Thermal Imaging Spectrometer).  
The observed values ranged between 40 and 160 J K$^{-1}$ m$^{-2}$ s$^{-1/2}$ \citep{Marshall2018AA}, 
with lower values ($<$\,50 J K$^{-1}$ m$^{-2}$ s$^{-1/2}$) being more common \citep{Tosi2019NatAs}. 
These findings are indicative of a highly porous structure with low thermal diffusivity \citep[][]{KellerEkkehard2020SSRv}.  
Thermal inertia, $I$ (J K$^{-1}$ m$^{-2}$ s$^{-1/2}$), is defined as 
\begin{equation}
I = \sqrt{k \rho_{b} c_{\rm p}}, 
\label{inertia}
\end{equation}
where 
$k$ is the thermal conductivity, 
$\rho_{b}$ = 538 \,kg\,m$^{-3}$ is the bulk density of the 67P nucleus  \citep{Martin2019}, 
$c_{\rm p}$\,=\,1000 J kg$^{-1}$ K$^{-1}$ is the assumed heat capacity of the pebbles \citep{Blum2017,Fulle2020MNRAS}.  
With $I$\,=\,40$-$160 J K$^{-1}$ m$^{-2}$ s$^{-1/2}$, 
we find $k$ = 3$\times$10$^{-3}$\,$-$\,5$\times$10$^{-2}$ W m$^{-1}$ K$^{-1}$.  
Thermal diffusivity is calculated by $\kappa = k / (\rho_{\rm b}\,c_{\rm p})$.
Substituting the above parameters, we obtain 
$\kappa$\,$\sim$\,10$^{-8}$$-$10$^{-7}$~m$^2$~s$^{-1}$.   
The conduction timescale for 289P is given by $\tau_{\rm c} \sim r_{\rm n}^2 / \kappa$, 
where $r_{\rm n}$$\sim$\,160\,m.  
Substitution $\kappa$ derives $\tau_{\rm c}$\,$\sim$\,8\,$\times$\,10$^{3-4}$\,yr.

\section{2013 Activity in comet 289P/Blanpain at $R_{\rm h}$ = 3.88\,au}
\label{2013Activity}

On  2013 July 5 and 6 (UT), observations by 
the Pan-STARRS and other telescopes 
confirmed the presence of a coma and tail for 
comet 289P/Blanpain at the heliocentric 
distance $R_{\rm h}$ = 3.88\,au, 
and the apparent visual magnitude was 
$m_{\rm V}$ = 19.5\,$\sim$\,17.5\,mag \citep{Williams2013CBET}.
Here, we estimate the relevant ejected dust mass, $M_{\rm d}^\prime$, 
focusing on the contribution from coma (not tail).     
The apparent visual magnitude, $m_{\rm V}$, was corrected to 
the absolute magnitude, $m_{\rm V}(1,1,0)$, using 
\begin{equation}
m_{\rm v} (1,1,0) = m_{\rm v} - 5\,{\rm log}(R_{\rm h}\,\Delta) + 2.5\,{\rm log}_{10} (\Phi_{\rm HM}(\alpha)), 
\label{absVmag}
\end{equation}
where $R_{\rm h}$ and $\Delta$ are the heliocentric 
and 
geocentric distances (both in au),  
and 
$\Phi_{\rm HM}(\alpha)$ is the Halley-Marcus composite phase 
function for dust scattering in coma 
\cite[][]{Schleicher1998Icar,Marcus2007ICQ_29_39M} in which  
$\alpha$ is the phase angle (in degrees).  
Substituting $R_{\rm h}$\,=\,3.88\,au, $\Delta$\,=\,2.87\,au, $\alpha$\,=\,2$^{\circ}$.5 
(i.e., $\Phi_{\rm HM}$(2$^{\circ}$.5)\,=\,0.90) 
(on UT 2013 July 6, 13:00 from NASA/JPL HORIZONS) 
and maximum $m_{\rm v}$\,=\,17.5\,mag into Equation~(\ref{absVmag}), 
we find the absolute magnitude $m_{\rm v} (1,1,0)$ = 12.2\,mag.  
The effective scattering cross-section of 
dust and nucleus within the photometry aperture, $C_{\rm e}$(km$^2$), 
is given by~\citep{Russell1916}
\begin{equation}
C_{\rm e}   = \frac{\pi \times 2.24 \times 10^{16}}{p_{\rm v}}10^{0.4({m_{\rm v}}_{\odot}- m_{\rm v}(1,1,0))}, 
\label{vis_Ce} 
\end{equation}
where ${m_{\rm v}}_{\odot}$ = $-$26.75\,mag is the 
apparent visual magnitude of the Sun \citep{Drilling2000}.
With $p_{\rm v}$=0.04 (Section~\ref{model}) and $m_{\rm v} (1,1,0)$ = 12.2\,mag into Equation~(\ref{vis_Ce}), 
we find $C_{\rm e}$ = 463\,km$^2$.   
By subtracting the nucleus cross-section ($C_{\rm n}$) 
from the effective scattering cross-section ($C_{\rm e}$),  
we obtain the cross-section of dust, $C_{\rm d}$ = 462.9\,$\pm$0.1\,km$^{2}$.    
Using Equation~(\ref{M_d}) with $a_{\rm d}$ = 0.5\,$\micron$ (visual wavelength),  
$\rho_{\rm d}$=1000\,kg\,m$^{-3}$ (assumed bulk density), 
and 
the obtained $C_{\rm d}$, 
we find the ejected dust mass $M_{\rm d}^\prime$ $\approx$ 3.1~$\times$~10$^{5}$\,kg.

The mechanism driving sudden activity near 4\,au remains uncertain. 
Crystallization of amorphous ice is unlikely 
due to the 289P's internal temperature of $\sim$\,180\,K (Section~\ref{W2Q}),  
exceeding the critical temperature of $\sim$140\,K \citep[][]{PrialnikJewitt2022}.  
Seasonal effects on the rotating nucleus could contribute to the distant activity, 
as previously unexposed subsurface ice-rich region in the dark hemisphere 
turned into sunlight.  
However, still unclear whether the resulting 
sublimation of water ice and venting of integrated internal gas pressure 
can fully explain the observed massive dust ejection.  
The Rosetta mission to 67P served as a precedent.  
Amorphous water ice is not yet confirmed 
within the nucleus \citep[][]{KellerEkkehard2020SSRv}. 
The outbursts from some ice patches \citep[][]{Filacchione2019SSRv}
and continuous (background) activity, including 
morphological changes on the bi-lobe shape
(e.g. mass-wasting, cliff collapse, pit formation) 
caused by diurnal and seasonal effects \citep{ElMaarry2019SSRv}, 
have been observed.  
The understanding of these physical 
processes remains limited \citep[][]{KellerEkkehard2020SSRv}. 
Future missions, such as 
a spacecraft sample-return from 289P in the 2030\,s \citep{Wakita2023LPI} 
and 
a flyby observation by Comet Interceptor (backup plan in 2035) \citep[][]{Schwamb2020,Jones2024SSRv}, 
would provide more understanding of the nature.




\bibliography{ref}
\bibliographystyle{aasjournal}




\begin{deluxetable*}{cccccccccccc}
\tablecaption{NEOWISE Observation Log \label{obslog}}
\tablewidth{0pt}
\tabletypesize{\scriptsize}
\tablehead{
\colhead{Visit}   &\colhead{\rm MJD\tablenotemark{a}}  & \colhead{UT Date}             & \colhead{RA\tablenotemark{b}}     & \colhead{DEC\tablenotemark{c}}    & \colhead{$R_{\rm h}$\tablenotemark{d}} & \colhead{$\Delta$\tablenotemark{e}} & \colhead{$\alpha$\tablenotemark{f}} & \colhead{$W1$\tablenotemark{g}} & \colhead{$W2$\tablenotemark{h}} \\
                  &                                &                                   &    (deg)                          &      (deg)                        &    (au)                               &          (au)                       &         (deg)                       &    (mag)                        &               (mag)           }
\startdata                                                                            
$A$               & 58786.0688                     &   2019-10-30 01:39:05.796         &   336.618                         &     -24.534                       & 1.2056	                            &           0.4018	                  &        49.5624                         & 15.181$\pm$0.563         	     &     \tablenotemark{i}	     \\
                  & 58786.1997                     &   2019-10-30 04:47:35.840         &   336.594                         &     -24.521                       & 1.2046                             &           0.4014	                  &        49.6867                         &   \tablenotemark{j}             &     \tablenotemark{j}	      \\
                  & 58786.2650                     &   2019-10-30 06:21:39.862         &   336.582                         &     -24.514                       & 1.2041	                            &           0.4012	                  &        49.7487                         &     \tablenotemark{i}           &      13.033$\pm$0.151         \\
                  & 58786.3304                     &   2019-10-30 07:55:54.884         &   336.570                         &     -24.508                       & 1.2035	                            &           0.4010			  &        49.8108                         & 15.800$\pm$0.392		     &      13.490$\pm$0.198         \\
                  & 58786.3959                     &   2019-10-30 09:30:09.910         &   336.558                         &     -24.501                       & 1.2030		                    &           0.4008                    &        49.8730	                   & 15.037$\pm$0.367		     &      \tablenotemark{j}           \\
                  & 58786.4613                     &   2019-10-30 11:04:24.932         &   336.547                         &     -24.494                       & 1.2025	                            &           0.4006		          &        49.9351	                   & 14.916$\pm$0.307	             &      12.231$\pm$0.077	     \\
                  & 58786.5267                     &   2019-10-30 12:38:28.954         &   336.535                         &     -24.487                       & 1.2020	                            &           0.4004	                  &        49.9970	                   &    \tablenotemark{j}            &      13.376$\pm$0.202         \\
                  & 58786.6576                     &   2019-10-30 15:46:59.002         &   336.512                         &     -24.474                       & 1.2009                             &           0.3999		          &        50.1212	                   & 15.565$\pm$0.259	             &      13.032$\pm$0.133        \\
                  & 58786.7884                     &   2019-10-30 18:55:18.046         &   336.489                         &     -24.460                       & 1.1999		                    &           0.3995	                  &        50.2452	                   &   \tablenotemark{j}             &      12.656$\pm$0.121        \\
\hline                                                                                
$B$               & 58859.9303                     &   2020-01-11 22:19:39.686         &   15.406                          &     +57.207                       & 1.0099	                            &           0.0910	 	          &        70.5400                         &    \tablenotemark{k}  	     &      \tablenotemark{k}           \\
                  & 58859.9957                     &   2020-01-11 23:53:54.721         &   15.721                          &     +57.442                       & 1.0102	                            &           0.0910			  &        70.3560                        &  12.471$\pm$0.045	             &       9.743$\pm$0.039	      \\
                  & 58860.1266                     &   2020-01-12 03:02:24.787         &   16.364                          &     +57.910                       & 1.0108	                            &           0.0910		          &        69.9862	                  &  12.529$\pm$0.052	             &       9.718$\pm$0.039	      \\
\enddata
\tablecomments{Visit~$A$~(MJD~58786: UT 2019-10-30, inbound) and Visit~$B$~(MJD~58860: UT 2020-01-11/12, outbound) are defined.}
\tablenotetext{a}{Modified Julian Date of the mid-point of the observation.}
\tablenotetext{b}{Right ascension (J2000).} 
\tablenotetext{c}{Declination (J2000).} 
\tablenotetext{d}{Heliocentric distance.} 
\tablenotetext{e}{{\it WISE}-centric distance.} 
\tablenotetext{f}{Phase angle.}
\tablenotetext{g}{Measured magnitude at $W1$ from the single-exposure frame.}
\tablenotetext{h}{Measured magnitude at $W2$ from the single-exposure frame.}
\tablenotetext{i}{SNR\,$<$\,1.}
\tablenotetext{j}{The 289P image is unconfirmed or too weak to measure the FWHM.}
\tablenotetext{k}{Contaminated by a star.}
\end{deluxetable*}

\clearpage

\begin{deluxetable*}{cccccccccc}
\tablecaption{Orbital Property of Comet 289P/Blanpain (2003~WY$_{25}$) \label{289P_orbit}}
\tablewidth{0pt}
\tablehead{
\colhead{$a$\tablenotemark{a}} & \colhead{$e$\tablenotemark{b}} & \colhead{$i$\tablenotemark{c}} & \colhead{$q$\tablenotemark{d}} & \colhead{$\omega$\tablenotemark{e}} & \colhead{$\Omega$\tablenotemark{f}} & \colhead{$Q$\tablenotemark{g}} & \colhead{$P_{\rm orb}$\tablenotemark{h}} & \colhead{$T_{\rm J}$\tablenotemark{i}}\\
            (au)                        &                                &      (deg)                     &    (au)               &          (deg)            &         (deg)             &    (au)    &    (yr)      }
\startdata
3.045                & 0.685            & 5.897               & 0.959                 & 9.849                   & 68.924    &   5.132        & 5.315 & 2.817 \\
\enddata
\tablecomments{ 
From NASA JPL Small-Body Database Lookup (2458746.5 (2019-Sep-20.0): Solution Date 2024-Jul-26)
}
\tablenotetext{a}{Semimajor axis.}
\tablenotetext{b}{Eccentricity.} 
\tablenotetext{c}{Inclination.} 
\tablenotetext{d}{Perihelion distance.} 
\tablenotetext{e}{Argument of perihelion.} 
\tablenotetext{f}{Longitude of ascending node.}
\tablenotetext{g}{Aphelion distance.}
\tablenotetext{h}{Orbital period.}
\tablenotetext{i}{Tisserand parameter with respect to Jupiter.  $T_J < $ 3.08 for comets and $T_J > 3.08$ for asteroids, if $a < a_J$ = 5.2\,au \citep{Jewitt_active15}.
  For reference, we list other comet-asteroid thresholds of $T_J = 3.05$ \citep{Tancredi2014Icar} and $T_J = 3.10$ \citep{HsiehHaghighipour2016Icar} \citep[see a review,][]{JewittHsieh2022}.}
\end{deluxetable*}

\clearpage

\begin{deluxetable}{ccccc}
\tablecaption{Magnitude and Flux Density \label{mag_flux}}
\tablewidth{0pt}
\tablehead{
       Visit      &     \colhead{$W1$\tablenotemark{a}} & \colhead{$W2$\tablenotemark{b}}  &     \colhead{$W1$\tablenotemark{c}}     & \colhead{$W2$\tablenotemark{d}}   \\
                  &       (mag)                          &       (mag)                        &       (Jy)                           &        (Jy)                           }
\startdata
  $A$\tablenotemark{e}    & 15.903$\pm$0.133                     & 13.035$\pm$0.066                   &  1.35$\pm$0.17$\times$10$^{-4}$       & 1.05$\pm$0.06$\times$10$^{-3}$        \\
  $B$\tablenotemark{f}   & 12.525$\pm$0.046                     &  9.714$\pm$0.039                   &  3.02$\pm$0.13$\times$10$^{-3}$       & 2.24$\pm$0.08$\times$10$^{-2}$        \\
\enddata
\tablecomments{Magnitudes are measured from the composite images and converted to flux densities for each band \citep{Wright2010AJ}.}
\tablenotetext{a}{Measured magnitude at $W1$ from the composite image.}
\tablenotetext{b}{Measured magnitude at $W2$ from the composite image.}
\tablenotetext{c}{Flux density, converted from magnitude at $W1$.}
\tablenotetext{d}{Flux density, converted from magnitude at $W2$.}
\tablenotetext{e}{MJD~58786: UT 2019-10-30, inbound. $R_{\rm h}$=1.20\,au, $\Delta$=0.40\,au, and $\alpha$=49.8$^{\circ}$.}
\tablenotetext{f}{MJD~58860: UT 2020-01-11/12, outbound. $R_{\rm h}$=1.01\,au, $\Delta$=0.09\,au, and $\alpha$=70.2$^{\circ}$.}
\end{deluxetable}

\clearpage

\begin{deluxetable*}{cccccc}
\tablecaption{ 
Weighted Mean of $W1$ and $W2$ Magnitudes
\label{lightcurve_log}}
\tablewidth{0pt}
\tabletypesize{\scriptsize}
\tablehead{
\colhead{Visit}   &\colhead{\rm MJD\tablenotemark{a}}   & \colhead{$W1$\tablenotemark{b}}   & \colhead{$W2$\tablenotemark{c}}       &  \colhead{$W_{\rm wm}$\tablenotemark{d}} \\
                  &                                     &    (mag)                          &               (mag)                   &             (mag)                       }
\startdata                                             
$A$               & 58786.0688                          & 15.181$\pm$0.563\tablenotemark{e} &     11.662\tablenotemark{h}	    &  12.056$\pm$0.188\\
                  & 58786.1997                          & 15.491\tablenotemark{f}           &     12.577\tablenotemark{h}	    &  13.474$\pm$0.166\\
                  & 58786.2650                          & 15.645\tablenotemark{f}           &     13.033$\pm$0.151\tablenotemark{g} &  13.561$\pm$0.135\\
                  & 58786.3304                          & 15.800$\pm$0.392\tablenotemark{e} &     13.490$\pm$0.198\tablenotemark{g} &  13.960$\pm$0.177\\
                  & 58786.3959                          & 15.037$\pm$0.367\tablenotemark{e} &     12.860\tablenotemark{h}           &  13.358$\pm$0.176\\
                  & 58786.4613                          & 14.916$\pm$0.307\tablenotemark{e} &     12.231$\pm$0.077\tablenotemark{g} &  12.390$\pm$0.075\\
                  & 58786.5267                          & 15.132\tablenotemark{f}           &     13.376$\pm$0.202\tablenotemark{g} &  13.924$\pm$0.168\\
                  & 58786.6576                          & 15.565$\pm$0.259\tablenotemark{e} &     13.032$\pm$0.133\tablenotemark{g} &  13.561$\pm$0.118\\
                  & 58786.7884                          & 15.997\tablenotemark{f}           &     12.656$\pm$0.121\tablenotemark{g} &  13.123$\pm$0.112\\
\enddata
\tablecomments{For light curve analysis with $W_{\rm wm}$ from Visit~$A$~(MJD~58786: UT 2019-10-30, inbound).}
\tablenotetext{a}{Modified Julian Date of the mid-point of the observation.}
\tablenotetext{b}{Magnitude at $W1$ from the single-exposure frame.}
\tablenotetext{c}{Magnitude at $W2$ from the single-exposure frame.}
\tablenotetext{d}{Weighted mean of $W1$ and $W2$ magnitudes.}
\tablenotetext{e}{Measured magnitude at $W1$ from Table~\ref{obslog}.}
\tablenotetext{f}{Interpolated/extrapolated magnitude at $W1$.  Assumed uncertainty is 0.3\,mag.}
\tablenotetext{g}{Measured magnitude at $W2$ from Table~\ref{obslog}.}
\tablenotetext{h}{Interpolated/extrapolated magnitude at $W2$.  Assumed uncertainty is 0.2\,mag.}
\end{deluxetable*}

\clearpage

\begin{deluxetable}{cccccc}
\tablecaption{Properties of Dust Production \label{final_results}}
\tablewidth{0pt}
\tablehead{
 Visit               &     \colhead{$D_{\rm e}$\tablenotemark{a}}  & \colhead{$C_{\rm d}$\tablenotemark{b}} & \colhead{$M_{\rm d}$\tablenotemark{c}} & \colhead{$Q_{\rm dust}$\tablenotemark{d}}       & \colhead{$Q_{\rm Af\rho}$\tablenotemark{e}}           \\
                     &         (km)                               &        (km$^2$)                       &       (kg)                            &       (kg\,s$^{-1}$)                           &  (kg\,s$^{-1}$)                     } 
\startdata
$A$\tablenotemark{f} &  1.43$\pm$0.04                             & 1.53$\pm$0.09                         &  4100$\pm$200                         &  1.0$\pm$0.1$\times$10$^{-2}$                  & 1.0$\pm$0.2$\times$10$^{-2}$\tablenotemark{h}        \\
$B$\tablenotemark{g} &  0.96$\pm$0.01                             & 0.64$\pm$0.08                         &  1700$\pm$200                         &  2.0$\pm$0.3$\times$10$^{-2}$                  & 4.0$\pm$0.2$\times$10$^{-2}$\tablenotemark{i}        \\
\enddata
\tablecomments{}
\tablenotetext{a}{Effective diameter of a circle having the same area as the sum of all the dust particles and nucleus.}
\tablenotetext{b}{Cross-section of all the dust particles.}
\tablenotetext{c}{Dust mass from Equation~(\ref{M_d}).}
\tablenotetext{d}{Dust production rate from Equation~(\ref{Qdust}).}
\tablenotetext{e}{Dust production rate from Equation~(\ref{Cremonese2020Qdust}).}
\tablenotetext{f}{MJD~58786: UT 2019-10-30, inbound. $R_{\rm h}$=1.20\,au, $\Delta$=0.40\,au, and $\alpha$=49.8$^{\circ}$.}
\tablenotetext{g}{MJD~58860: UT 2020-01-11/12, outbound. $R_{\rm h}$=1.01\,au, $\Delta$=0.09\,au, and $\alpha$=70.2$^{\circ}$.}
\tablenotetext{h}{$Af\rho$ = 2.9$\pm$0.4\,cm from Equation~(\ref{Afro}).}
\tablenotetext{i}{$Af\rho$ = 10.6$\pm$0.4\,cm from Equation~(\ref{Afro}).}
\end{deluxetable}

\clearpage


\clearpage

\clearpage

\clearpage

\clearpage

\clearpage

\clearpage

\clearpage
\centering

\clearpage
\centering

\clearpage

\clearpage

\clearpage

\clearpage

\clearpage

\clearpage

\end{document}